\documentclass[twocolumn]{autart}    
\usepackage{graphics} 
\usepackage{epsfig} 
\usepackage{amsmath} 
\usepackage{amssymb}  
\usepackage{cite}
\usepackage{natbib}
\usepackage{bm}
\usepackage{subfigure}
\usepackage{acronym}
\usepackage{paralist}
\usepackage{multicol}
\usepackage{epstopdf}
\usepackage{placeins}
\usepackage{color}
\usepackage{url}
\usepackage{scalerel}
\usepackage{hyperref}
\usepackage{calligra}
\DeclareMathAlphabet{\mathcalligra}{T1}{calligra}{m}{n}
\DeclareFontShape{T1}{calligra}{m}{n}{<->s*[2.2]callig15}{}
\newcommand{\sr}{\ensuremath{\mathcalligra{r}}}
\newcommand{\vectorproj}[2][]{\textit{proj}_{\vect{#1}}\vect{#2}}
\newcommand{\vect}{\mathit}

\DeclareMathOperator{\atan2}{atan2}
\DeclareMathOperator\artanh{artanh}

\DeclareMathOperator*{\argmin}{\arg\!\min}
\DeclareMathOperator*{\argmax}{\arg\!\max}
\usepackage{mathtools}
\newcounter{mytempeqncnt}
\newtheorem{Lemma}{Lemma}
\newtheorem{Lemma1}[Lemma]{Lemma}

\newtheorem{Theorem}{Theorem}
\newtheorem{Theorem_interception}[Theorem]{Theorem}
\newtheorem{Theorem_window}[Theorem]{Theorem}
\newtheorem{Theorem_avoidance}[Theorem]{Theorem}
\newtheorem{Remark}{Remark}
\newtheorem{Remark1}[Remark]{Remark}
\newtheorem{Remark2}[Remark]{Remark}

\newtheorem{Definition}{Definition}
\newtheorem{Definition1}[Definition]{Definition}
\newtheorem{Definition2}[Definition]{Definition}
\begin{document}

\begin{frontmatter}

\title{A Hybrid Approach to Persistent Coverage in Stochastic Environments\thanksref{footnoteinfo}} 

\thanks[footnoteinfo]{The authors would like to acknowledge the support of the Automotive Research Center (ARC) in accordance with Cooperative Agreement W56HZV-14-2-0001 U.S. Army TARDEC in Warren, MI and the support by an Early Career Faculty grant from NASA’s Space Technology Research Grants Program. The material in this paper was partially presented at the 56th IEEE Conference on Decision and Control, December 12-15, 2017, Melbourne, Australia and the 2018 American Control Conference, June 27-29, 2018, Milwaukee, Wisconsin. See Section 1.2 for a comparison between the present work and the conference versions.}

\author[Ann Arbor]{William Bentz}\ead{wbentz@umich.edu},
\author[Ann Arbor]{Dimitra Panagou}\ead{dpanagou@umich.edu}

\address[Ann Arbor]{Department of Aerospace Engineering, University of Michigan, 1320 Beal Ave, Ann Arbor, MI, 48109, USA}

\begin{keyword}                           
Coverage Control; Collision avoidance; Autonomous mobile robots; Cooperative control; Multi-agent systems .               
\end{keyword}                             

\begin{abstract}                          
This paper considers the persistent coverage of a 2-D manifold that has been embedded in 3-D space. The manifold is subject to continual impact by intruders which travel at constant velocities along arbitrarily oriented straight-line trajectories. The trajectories of intruders are estimated online with an extended Kalman filter and their predicted impact points contribute normally distributed decay terms to the coverage level. A formal hybrid control strategy is presented that allows for power-constrained 3-D free-flyer agents to persistently monitor the domain, track and intercept intruders, and periodically deploy from and return to a single charging station on the manifold. Guarantees on intruder interception with respect to agent power lifespans are formally proven. The efficacy of the algorithm is demonstrated through simulation.
\end{abstract}

\end{frontmatter}\vspace{-2mm}
\section{Introduction}\vspace{-2mm}
\subsection{Background}\vspace{-2mm}
The advent of inexpensive autonomous research platforms has spurred recent interest in teams of mobile sensors collaborating on complex surveillance and monitoring tasks. Coverage control problems have been particularly popular due to their numerous applications: e.g., environmental monitoring \citep{smith2011persistent}, battlefield surveillance \citep{bokareva2006wireless}, lawn mowing and vacuuming, search and rescue \citep{murphy2008search}, and hull inspections \citep{choset1999path,hollinger2013active}. The latter application is actively supported by NASA whose work on the Mini AERCam paves the way for a future of extravehicular robotic (EVR) free flyers performing independent visual inspections of spacecraft exterior areas of interest \citep{fredrickson2004application}. Free flyer visual inspection is the primary motivating example for our work.
\vspace{-2mm}

Coverage is often partitioned into three classes of problems: static, dynamic, and persistent. Static coverage problems (e.g., area coverage, k-coverage and point coverage) often explore the optimal arrangement of sensor nodes in a network and the agents tend to immobilize after this arrangement has been achieved \citep{Cortes_TRA2004}. Dynamic coverage problems involve the active exploration of a domain. Agents typically must sweep their sensors over all points of a domain until some desired level of coverage has been achieved \citep{Hussein_CST07,liu2013dynamic,Stipanovic_SafeCoverage2012}. Persistent coverage is often similar to dynamic coverage with the addition of information decay within the environment: i.e., agents are required to continually return to areas of interest in order to restore a deteriorating coverage level.
\vspace{-2mm}

The term "persistent coverage" appears as early as \citet{Hokayem_CDC07} where agents must cover all points in a 2-D convex polygonal domain every $T^{\star}$ time units. This was accomplished with the design of concentric polygonal trajectories with agents following closed paths in steady state. The work in \citet{song2013persistent} is similar but also introduces a linear coverage decay rate for specific points of interest. In this paper, as well as \citet{smith2012persistent}, controller design is akin to regulating the velocity along paths generated offline to increase observation time at select points of interest. As the decay rates are known and time invariant, optimal speed control is computed via linear programming.\vspace{-2mm}

Palacios-Gas\'{o}s et. al have published multiple works recently on persistent coverage \citep{palacios2016multia,palacios2016multib,palacios2017optimal} which build specifically upon \citet{smith2012persistent}. While the earlier work assumed both the existence and knowledge of an optimal path to cover all points of interest, \citet{palacios2016multia} uses techniques from discrete optimization and linear programming to iteratively compute this path. The effect is that if the coverage decay rate of a specific point of interest is found to be insufficient to justify the transit time required to service it, then the point may be removed online from the path of the robot. Prior works, i.e., \citet{smith2012persistent,song2013persistent}, would have instead driven the robot to quickly pass through the point. Similar techniques are used in \citet{mitchell2015multi} which also considers that agents must periodically return to refueling depots.\vspace{-2mm}

In \citet{Hubel2008} and \citet{Song20112749}, the desired coverage level of the domain is maintained with density maps that yield additional observation time at select areas of interest. In \citet{Song20112749} the maps are time-invariant while \citet{Hubel2008} considers time-varying density maps that may be designed around moving points of interest (e.g., aerial surveillance targets). However, the latter work only uses density maps in the derivation of control laws and not in the differential equations governing coverage level evolution.\vspace{-2mm}

Common themes through all of these persistent coverage works are convex 2-D domains, predictable environments, and simplified sensing and dynamic models for agents. Coverage surfaces embedded in $\mathbb{R}^3$ are considered in \citet{ChengKumar2008}; however, this work is closer to that of \citet{Hokayem_CDC07} in that agents also follow preplanned trajectories without considering spatially-dependent coverage decay maps.\vspace{-2mm}

Monitoring of stochastic environments is presented in \citet{yu2015persistent,pasqualetti2014camera}, outside of the strict persistent coverage formulation. In \citet{yu2015persistent}, the authors consider that agents must observe events at multiple points of interest and the precise arrival times of events are unknown \textit{a priori}. Arrival time statistics are used to inform a multi-objective scheduling protocol that results in fixed cyclic servicing policies. In \citet{pasqualetti2014camera}, the environment contains smart intruders which actively attempt to evade a camera surveillance network. Camera motion is restricted to a single pan axis and thus the the system model is essentially that of a 1-D pursuit evasion problem.\vspace{-2mm}
\subsection{Contribution}\vspace{-2mm}
In this paper, we present (i) a formal hybrid control strategy for multi-agent persistent coverage of non-planar convex surfaces embedded in $\mathbb{R}^3$ that does not make overly simplifying assumptions with respect to agent dynamic and sensing models, (ii) guarantees on agent interception of stochastic intruders, and (iii) an energy-aware agent deployment and scheduling protocol. To the best of our knowledge, we are the first to present a formal hybrid approach to persistent coverage.\vspace{-2mm}

Although many of the cited works use density functions to encode points of interest, the difference in our approach is both subtle and powerful. The density function in \citet{Hubel2008} evolves subject to the motion of an intruder and informs the control laws; however, it has no effect upon the dynamics of the coverage level. Thus, only the intruder's current location has any influence on the motion of the agents, and the time-history of the intruder's trajectory is forgotten. This necessitates that agents must travel faster than targets in order to cover points associated with peaks in the density function before they vanish. Works that do include a density function in the coverage level evolution, e.g \citet{palacios2016multia,palacios2016multib,palacios2017optimal}, tend to have fixed decay rates that cannot respond or adapt to a changing environment. In contrast, our algorithm utilizes a time-varying density function, which is estimated online via extended Kalman Filter, to directly encode coverage decay over the surface around the predicted impact points of intruders. This encodes a memory effect which drives some agents to follow coverage gradients towards areas that have previously been or will soon be impacted by intruders.\vspace{-2mm}

Our agents operate with finite resources and are required to periodically return to a refueling station while observing stochastic events at locations and times that are not known \textit{a priori}. This approach is different from related works, such as \citet{yu2015persistent} and \citet{mitchell2015multi}, where the locations of events are fixed and the authors are concerned with optimal servicing routes between these known stations.\vspace{-2mm}

This hybrid system is a successor to our previous works in \citet{Bentz_CDC_2017, Bentz_ACC_2018}. In \citet{Bentz_CDC_2017}, we derived the first of our hybrid modes (i.e., local coverage mode) and our intruder state estimator. However, the agents had no power constraints and the approach was unable to provide any formal guarantees on intruder interception without additional operating modes. In \citet{Bentz_ACC_2018}, we derived these additional modes to present a hybrid approach to persistent coverage. Agents were now scheduled to intercept intruders and followed path-length optimal trajectories. However, formal guarantees on intruder interception were still limited to cases in which no collision-avoidance deadlocks had occurred. Furthermore, agents did not make effective use of local coverage mode as they would often travel to the predicted impact points of intruders and then remain stationary until the moment of impact thus contributing to a rising coverage error.\vspace{-2mm}

This work extends the interception guarantee of \citet{Bentz_ACC_2018} to an arbitrary number of collision avoidance maneuvers and presents an entirely new method of collision avoidance over the prior works. It also reformulates numerous guard conditions within the automaton to allow agents to explore actively around the predicted impact points of intruders. Furthermore, this contribution revises our sensing function definition utilized in the prior works which suffered from a singularity at the sensing cone vertex.\vspace{-2mm}

This paper is organized as follows: Section \ref{Problem Formulation} describes the agents sensing and kinematic models and provides an overview of our hybrid control strategy, Section \ref{Information Decay} presents the trajectory estimator for particle intruders and defines our coverage decay rate map, Sections 3-6 describe each hybrid mode in detail, Section 7 verifies the algorithm in simulation, Section 8 summarizes our contributions and Section 9 is an appendix containing the formal definition of our hybrid automaton. 

\section{Problem Formulation}\label{Problem Formulation}\vspace{-3mm}
\subsection{Agent Modeling}\vspace{-3mm}
Consider a network of spherical autonomous agents indexed $i \in \{1,...,N\}$, of radius $\sr_i$, whose motion is subject to 3-D rigid body kinematics \citep{Beard}:\vspace{-3mm}
 \begin{multline} \label{eq:1}
\scriptsize 
 \begin{bmatrix}
 \dot{x_{i}} \\
 \dot{y_{i}} \\
 \dot{z_{i}} \\
 \end{bmatrix}
 =
 \left[\begin{matrix}
 \cos \Theta_{i}\cos \Psi_{i} & \sin \Phi_{i}\sin \Theta_{i}\cos \Psi_{i}-\cos \Phi_{i}\sin \Psi_{i} \\ 
 \cos \Theta_{i}\sin \Psi_{i} & \sin \Phi_{i}\sin \Theta_{i}\sin \Psi_{i}+\cos \Phi_{i}\cos \Psi_{i} \\  
   -\sin \Theta_{i} & \sin \Phi_{i}\cos \Theta_{i}\end{matrix}\right. \\
\scriptsize \left.\begin{matrix}
\cos \Phi_{i}\sin \Theta_{i}\cos \Psi_{i}+\sin \Phi_{i}\sin \Psi_{i}\\
\cos \Phi_{i}\sin \Theta_{i}\sin \Psi_{i}-\sin \Phi_{i}\cos \Psi_{i}\\
\cos \Phi_{i}\cos \Theta_{i}\end{matrix}\right]
\begin{bmatrix}
u_{i}\\
v_{i}\\
w_{i}
\end{bmatrix},
\end{multline}
\begin{equation} \label{eq:2}
\scriptsize
\begin{bmatrix}
\dot{\Phi_{i}}\\
\dot{\Theta_{i}}\\
\dot{\Psi_{i}}\\
\end{bmatrix}
=
\begin{bmatrix}
1 & \sin \Phi_{i} \tan \Theta_{i} & \cos \Phi_{i} \tan \Theta_{i} \\

0 & \cos \Phi_{i} & -\sin \Phi_{i} \\
0 & \sin \Phi_{i} \sec \Theta_{i} & \cos \Phi_{i} \sec \Theta_{i} 

\end{bmatrix}
\begin{bmatrix}
q_{i}\\r_{i}\\s_{i}\\
\end{bmatrix},\normalsize
\end{equation} where $p_{i}=\left[x_{i}\;\;y_{i}\;\;z_{i}\right]^{T}$ is the position vector and $\Omega_{i}=\left[\Phi_{i}\;\;\Theta_{i}\;\;\Psi_{i} \right]^{T}$ is the vector of 3-2-1 Euler angles taken with respect to a global Cartesian coordinate frame $\mathcal{G}$ with origin $\mathcal{O}$. The linear velocities $\left[u_{i}\;\;  v_{i}\;\;w_{i}\right]^{T}$ and angular velocities $\left[q_{i}\;\;  r_{i}\;\;s_{i}\right]^{T}$ are both presented in the body fixed frame $\mathcal{B}_i$ with origin $p_i$. The state vector of agent $i$ is defined as $\tilde{q}_{i}=[p_{i}^{T}\;\;\Omega_{i}^{T}]^{T}$. In the sequel, the rotation matrices of \eqref{eq:1} and \eqref{eq:2} shall be denoted $\mathcal{R}_1$ and $\mathcal{R}_2$ respectively. The agents travel within a stationary domain, $\mathcal{D} \subset \mathbb{R}^{3}$. Their task is to survey a two-dimensional manifold, $\mathcal{C} \subset \mathcal{D}$, known as our surface of interest. For the purpose of this work we assume that the surface is an ellipsoid of revolution; however, it should be noted that the coverage laws, as well as the collision avoidance strategy, can be easily adapted for any convex surface. The ellipsoid has semi-major axis $x_{\mathcal{C},r}$ and semi-minor axis $z_{\mathcal{C},r}$ aligned with the global coordinate axes $\hat{x}_{\mathcal{G}}$ and $\hat{z}_{\mathcal{G}}$ respectively with center at $\mathcal{O}$. The circumflex (i.e., hat) symbols denote unit vectors.

\begin{figure}[h]
\centering
\includegraphics[width=0.9\columnwidth,clip]{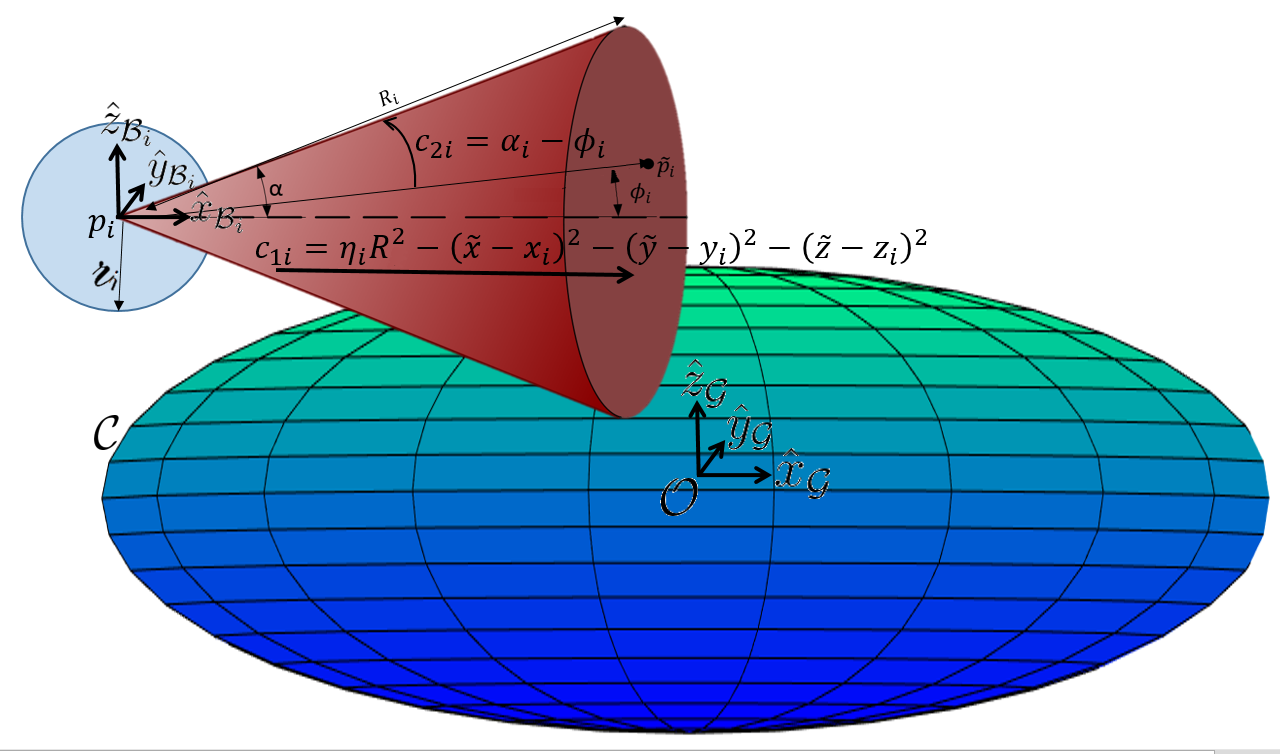}
\caption{Agent $i$ is modeled as a sphere of radius $\sr_i$ and has a forward facing sensor footprint, $\mathcal{S}_i$. Sensing constraint functions $c_{k i},\,\forall k \in \{1,2\}$, encode a decay in sensing quality along the depth and towards the periphery of $\mathcal{S}_i$.}
\label{fig:space_model}
\end{figure}

Each agent, $i$, is equipped with a forward facing sensor whose spherical sector footprint shall be referred to as $\mathcal S_i$. This model, though intended to be generic, is similar to conical camera models presented in other works on dynamic coverage (see \citet{Xie2013}). Our model differs in terms of its heterogeneity, i.e. $\mathcal S_i$ provides anisotropic sensing data that degrade in quality towards the periphery of the footprint and changes with respect to distance from the sensor. Degradation over distance is not monotonically decreasing but instead contains a peak located near the vertex of $\mathcal{S}_i$ as in \citet{hexsel2013distributed}. This is motivated by the fact that the probability of event detection by a camera decreases when either very far from or very close to the lens. Anisotropic sensing is encoded through the definition of the sensing constraint functions for each agent $i$:\vspace{-3mm}
\begin{subequations}
\label{coverage_functionals}
\begin{align} \label{eq:3}
c_{1i} &= \beta_iR^{2}-(\tilde{x}-x_{i})^{2}-(\tilde{y}-y_{i})^{2}-(\tilde{z}-z_{i})^{2}, \\
\label{eq:4}
c_{2i} &= \alpha_{i}-\phi_{i},
\end{align}
\end{subequations} for $\beta_i=\min\{1,\eta_i\left(\left(\tilde{x}-x_i\right)^2+\left(\tilde{y}-y_i\right)^2-\left(\tilde{z}-z_i\right)^2\right)\}$ with real constant $\eta_i>>1$. $R$ is the sensing range, $\tilde{p}_{i}=\left[\tilde{x}\;\;\tilde{y}\;\;\tilde{z}\right]^{T}$ is the position of a point within $\mathcal S_i$ with respect to $\mathcal{G}$, $\alpha_{i}$ is the angle between the periphery and centerline of the spherical sector (the $\hat{x}_{\mathcal B_i}$ axis), and $\phi_{i}$ is the angle between $r_{\tilde{p}_i/p_i}=\tilde{p}_{i}-p_i$ (resolved in $\mathcal{G}$ by construction) and the $\hat{x}_{\mathcal B_i}$ axis given as the inverse cosine of the dot product of $\hat{r}_{\tilde{p}_i/p_i}$ and $\hat{x}_{\mathcal{B}_i}$ resolved in $\mathcal{G}$: $ \phi_i= \arccos \left(\hat{r}_{\tilde{p}_i/p_i} \cdot \hat{x}_{\mathcal{B}_i} |_{\mathcal{G}} \right)$. Note that: $\hat{r}_{\tilde{p}_i/p_i}=\frac{1}{\sqrt{(\tilde{x}-x_{i})^{2}+(\tilde{y}-y_{i})^{2}+(\tilde{z}-z_{i})^{2}}}
\left[
(\tilde{x}-x_{i}) \; \;
(\tilde{y}-y_{i}) \; \;
(\tilde{z}-z_{i})
\right]^T$, and $\hat{x}_{\mathcal{B}_i} |_{\mathcal{G}}$ is determined by multiplying $\mathcal{R}_1$ by $\left[1\;\;0\;\;0\right]^{T}$: $\hat{x}_{\mathcal{B}_i} |_{\mathcal{G}}=\begin{bmatrix}
\cos \Psi_i \cos \Theta_i \; \;
\sin \Psi_i \cos \Theta_i \; \;
-\sin \Theta_i
\end{bmatrix}^T$. Agent $i$ is thus capable of detecting objects that lie within an angle of $2 \alpha_i >0$ about the $\hat{x}_{\mathcal B_i}$ axis and a range of $R>0$. The model for agent $i$ is depicted in Fig. \ref{fig:space_model}.

Let us denote $\max \{0,c_{ki}\}=C_{ki}$. One can define the sensing function that represents the quality of information available at each point over the sensing domain as:\vspace{-3mm}
\begin{align} \label{eq:7}
S_{i}(\tilde{q}_{i},\tilde{p})=
\begin{cases}
\hfill \frac{C_{1i}C_{2i}}{C_{1i}+C_{2i}}, \hfill &  \text{if $card\left(\bar{C}_i\right)<2 \land r_{\tilde{p}_i/p_i}>0$;}\\
\hfill 0, \hfill & \text{otherwise,}
\end{cases}
\end{align} where $\bar{C}_i$ is the set of zero elements in $C_{ki}$. $S_{i}(\tilde{q}_{i},\tilde{p})$ takes a value of zero outside of $\mathcal{S}_i$. Note that $S_{i}(\tilde{q}_{i},\tilde{p})$ is defined over all of $\mathcal{D}$ and thus has static bounds. $S_{i}(\tilde{q}_{i},\tilde{p})$ is continuous in $\tilde{p}$ while taking a value of zero along $\partial \mathcal{S}_i$. In verifying this continuity, it is important to note that $S_{i}(\tilde{q}_{i},\tilde{p})$ approaches zero from within $\mathcal{S}_i$ in the limit that either $card\left(\bar{C}_i\right)=2$ or $r_{\tilde{p}_i/p_i}=0$ are satisfied. The former condition may be verified by taking a limit of the first piecewise definition of \eqref{eq:7} as $C_{1i}$ and $C_{2i}$ tend to zero. The latter condition results from our definitions of $\beta_i$ and $\eta_i$ which encode that $S_{i}(\tilde{q}_{i},\tilde{p})$ drops off rapidly when in very close proximity to the vertex of $\mathcal{S}_i$. Increasing $\eta_i>>1$ has the effect of shifting the sensing drop off point closer to $p_i$. Define the coverage level provided by agent $i$ at time $t$ as:
\vspace{-3mm}
\begin{align}
\label{Agent i coverage level}
Q_i(t,\tilde p)=\int_0^t S_i\left(\tilde{q}_i(\tau),\tilde p\right)C\left(\tilde{p}\right)\,d\tau,
\end{align} where $C$ is defined as: $C\left(\tilde{p}\right)=
\begin{cases}
1,    \hfill & \text{ $\forall \tilde{p} \in \mathcal{C}$;} \\
\hfill 0, \hfill & \text{$\forall \tilde{p} \notin \mathcal{C}$,}
\end{cases}$ and encodes that the accumulation of sensing information only occurs along our surface of interest, $\mathcal{C}$.\vspace{-2mm}

As the agents cover $\mathcal{C}$, a set of $N_p$ high-speed particle intruders denoted $k \in \{1,...,N_p\}$, each of which travels in an arbitrary direction at constant velocity, pass through the domain. The particles are assumed to be uncontrolled and cannot deviate from their initial trajectories. No assumptions are made with respect to the source of the particles or whether they are intelligently generated. Each particle shall have an associated map decay term, $\Lambda_k\left(\tau,\tilde{p}\right)$, which is defined later in Section \ref{Decay Rate Map Evolution}. We may now define the global coverage level:
\vspace{-3mm}
\begin{align}
\label{global coverage function}
Q(t,\tilde p)=\sum_{i=1}^{N}Q_i(t,\tilde p)-\sum_{k=1}^{N_p}\int_0^t\Lambda_k\left(\tau,\tilde{p}\right)C\left(\tilde{p}\right)\,d\tau.
\end{align} In this work, coverage refers to the accumulation of sensing data over time. Points, $\tilde{p}$, are said to be sufficiently covered when $Q\left(t,\tilde{p}\right)\geq C^{\star}$. The goal is to derive a hybrid control strategy which persistently sweeps $\mathcal{S}_i$ across $\mathcal{C}$ while emphasizing surveillance within some bound of the predicted impact points of particles $k \in \{1,...,N_p\}$ on $\mathcal{C}$. More specifically, we establish theoretical guarantees on the worst case path length from any agent to any arbitrary impact point thus guaranteeing interception for prescribed bounds on intruder speed, detection range, and agent velocity. This must be done while avoiding collisions. Let us define collision and interception.\vspace{-2mm}
\begin{Definition2}
\label{Definition2}
\textnormal{Agent $i$ is said to have intercepted particle $k$ if $i$ is within a $\varepsilon_1$ bound of the estimated impact point of $k$ for a finite interval of time leading up to the impact. Agent $i$ shall spend this duration of time sweeping the area in local coverage thus gathering information.
}
\end{Definition2}\vspace{-2mm}
Note that intruders are unaffected by agents and shall always impact the surface and then disappear. This does not damage the agent which is free to resume other tasks upon conclusion of interception.
\begin{Definition1}
\label{Definition1}
\textnormal{Agent $i$ avoids collision so long as $\lVert p_i(t)-p_j(t) \rVert > \sr_i+\sr_j, \; \forall t \geq 0, \; \forall j\neq i \in \{1,...,N\}$ and $\lVert n_i\rVert>\sr_i$ where the vector $n_i$ has direction normal to $\mathcal{C}$ and length equal to the Euclidean distance of its intersection point on $\mathcal{C}$ to $p_i$.
}
\end{Definition1}\vspace{-2mm}
Agents operate with finite power resources and are required to return every $T^{\star}$ time units to a fueling station denoted $\mathcal{F}$. Thus, a scheduling protocol is derived whereby agents periodically deploy from $\mathcal{F}$ to cover within assigned partitions of $\mathcal{C}$. These partitions are bounded by latitude lines and are sorted by geodesic distance from $\mathcal{F}$ with agents deploying to the partition furthest from $\mathcal{F}$ and then transferring between adjacent partitions every $\frac{T^\star}{N}$ time units as their power resource dwindles requiring a return to $\mathcal{F}$ within $T^\star$ time units after deployment. This partitioning scheme also has the benefit of ensuring that the network of agents is well distributed across $\mathcal{C}$ with agents nominally assigned to intercept intruders with predicted impact points within their own partition.

\vspace{-5mm}Agent $i$ is capable of localizing itself in $\mathcal{G}$ and detecting whether there exists $j$ such that $\lVert p_i(t)-p_j(t) \rVert \leq R$. Furthermore, agents $i$ and $j$ can communicate their deployment times to one another. A centralized network is required to publish the current coverage level $Q(t,\tilde p)$\footnote{In practice, it is not necessary to publish $Q(t,\tilde p), \forall \tilde{p} \in \mathcal{D}$ to every agent. For agent $i$ to compute its local coverage control signal, it need only values for $Q(t,\tilde{p})$ within a closed ball of radius $R$ due to the fact that $S_i(\tilde{q}_i,\tilde{p})=0, \forall \tilde{p} \notin \bar{B}_Rp_i(t)$. This substantially reduces the communication overhead.} to all agents and to estimate the trajectories of intruders using an omnidirectional range sensor whose measurements are fed through an extended Kalman filter. Computation of $Q(t,\tilde p)$ is contingent upon continuous transmission of agent state $\tilde{q}_i$ to the centralized network. The centralized network assigns each intruder to an unassigned agent at closest latitude to the predicted impact point. It also transmits detection time as well as estimated location and time of impact to the agent.\vspace{-3mm}
\subsection{Intruder Modeling}\label{Information Decay}\vspace{-3mm}
We assume that the omnidirectional range sensor (e.g., LiDAR) is co-located with $\mathcal{O}$ and provides measurements of each particle's position in spherical coordinates. We also assume that particle detection and state estimate initialization occur while the distance of the particle from $\mathcal{O}$ is greater than or equal to $R_{det}+x_{\mathcal{C},r}$ where $R_{det}$ is a lower bound on distance from detection to impact. We define the model for the motion of particle $k$:
\vspace{-3mm}
\begin{align} 
\dot{\tilde{q}}_k\left(t\right)&= \left[\begin{matrix}
0_{3\times3} & \mathbb{I}_{3\times3} \\
0_{3\times3} & 0_{3\times3}
\end{matrix}\right]\tilde{q}_k\left(t\right), \label{Markov State} \\
\tilde{z}_k\left(t\right)
&=
\begin{bmatrix}
\sqrt{x_k^2+y_k^2+z_k^2} \\
\atan2{\left(y_k,\,x_k\right)} \\
\arccos{\left(\frac{z_k}{\sqrt{x_k^2+y_k^2+z_k^2}}\right)}
\end{bmatrix}+\epsilon, \label{Markov Output}
\end{align}
where $\tilde{q}_k=\left[x_k,\,y_k,\,z_k,\,\dot{x}_k,\,\dot{y}_k,\,\dot{z}_k\right]^{T}$ and $\tilde{z}_k=\left[\rho_k,\,\theta_k,\,\psi_k\right]^{T}$ are the Cartesian state and spherical coordinate measurement vectors of particle $k$ resolved in $\mathcal{G}$. We assume that particle speed is upper bounded such that $\sqrt{\dot{x}_k^2+\dot{y}_k^2+\dot{z}_k^2}\leq U_{max}^{int}$. $\rho_k$, $\theta_k$, and $\psi_k$ are the range, azimuthal angle, and polar angle of $k$ respectively. In the sequel, the matrix in \eqref{Markov Output} shall be denoted $\tilde{h}\left(x_k,\,y_k,\,z_k\right)$. Assume that the measurement noise, $\epsilon$, is zero-mean Gaussian and has covariance $\mathbf{R}=diag\left(\sigma_\rho^2,\sigma_\theta^2,\sigma_\psi^2\right)$. This system models high-speed particles incident upon a surface with negligible drag (e.g., micrometeoroids impacting a spacecraft hull); thus, it is reasonable to omit the process noise. The state and covariance estimates, $\hat{\tilde{q}}$ and $\mathbf{P}_k$, are computed with a continuous-time extended Kalman filter which is initialized upon particle $k$'s detection at time $t_{dk}$.\vspace{-3mm}
\subsection{Information Decay}\label{Decay Rate Map Evolution}\vspace{-3mm}
At any time $t$, we define our decay rate map for particle $k$ in terms of its predicted position and covariance evolution over a horizon $T_{H,k}(t)$. As the particles are assumed to travel at fixed velocities\footnote{The guarantee of intruder interception presented in this work can be extended to intruders with time-varying velocities that are bounded by $U^{int}_{max}$. However, it is still required that intruders follow straight line trajectories such that the network may estimate fixed impact points.}, the predicted values for Cartesian position $\tilde{p}'_k\left(t+\tau\right)$ and associated covariance $\mathbf{P}_k\left(t+\tau\right)$ are defined as
$\tilde{p}'_k\left(t+\tau\right)=G\left(\tau\right)\hat{\tilde{q}}_k\left(t\right)$, and $\mathbf{P'}_k\left(t+\tau\right)=G\left(\tau\right)\mathbf{P}_k
\left(t\right)G\left(\tau\right)^{T}$ respectively where $G\left(\tau\right)=\left[\mathbb{I}_{3\times3} \;\; \tau \mathbb{I}_{3\times3}\right]$ and $\hat{\tilde{q}}_k\left(t\right)$ is our current estimate for $\tilde{q}_k\left(t\right)$. We define the decay rate map associated with particle $k$ as the integral of our predicted normal distribution $\mathcal{N}\left(\tilde{p}'_k\left(t+\tau\right),\mathbf{P'}_k\left(t+\tau\right)\right)$ through horizon $T_{H,k}$:\vspace{-10mm}
\small
\begin{align}
\label{Decay Map}
\Lambda_k\left(t,\tilde{p}\right)=\int_0^{T_{H,k}(t)}\lambda_k\mathcal{N}\bigl(\tilde{p}'_k\left(t+\tau\right),\mathbf{P'}_k\left(t+\tau\right)\bigr)\,d\tau,
\end{align}
\normalsize
where $\lambda_k>0$ is a tuning parameter for the decay rate. We recommend choosing $\lambda_k<1$ as this typically allows for the rate of coverage to exceed the decay rate over points intersecting $\mathcal{S}_i$. For $t<t_{dk}$, define $\Lambda_k\left(t,\tilde{p}\right)=0,\,\forall \tilde{p} \in \mathcal{D}$. Our formulation for \eqref{Decay Map} essentially takes a normal distribution for the position of particle $k$ at time $t$ and cumulatively propagates it forward in time up to our horizon $T_{H,k}(t)$. The horizon is lower-bounded by an estimate of the remaining time until impact of particle $k$ on $\mathcal{C}$. This may be computed using $\tilde{q}_k\left(t\right)$ along with the surface geometry. With this design, $Q(t,\tilde p)$ decays along the predicted trajectory of $k$ with tapering omnidirectional decay rates spreading out from the predicted path. This design lends itself naturally to our local coverage formulation, which is gradient following in nature, in that the agents may follow these tapered decay paths towards the predicted impact points on our surface of interest.

\subsection{Algorithmic Overview}
\begin{figure}[h]
\centering
\includegraphics[width=0.9\columnwidth,clip]{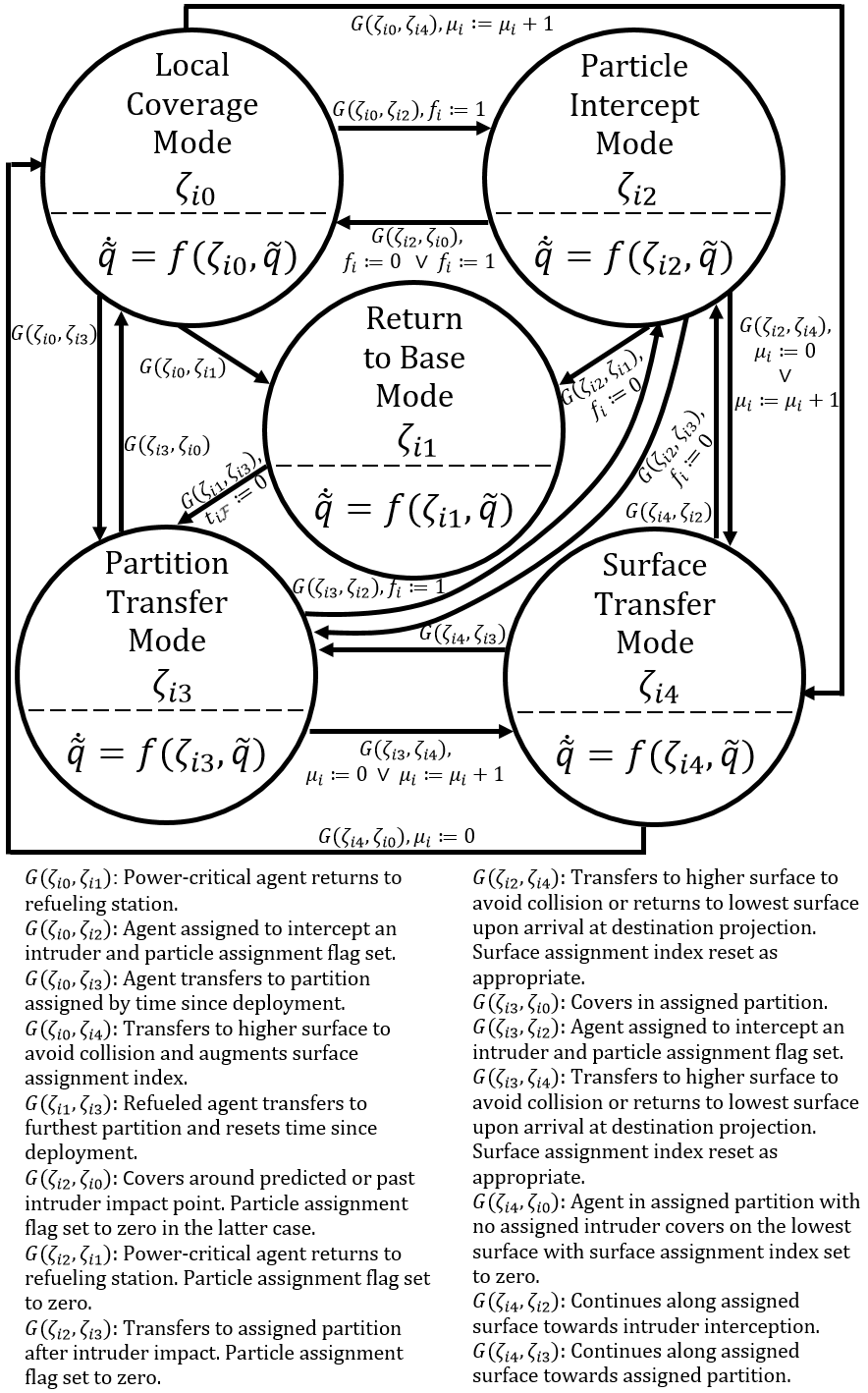}
\caption{Agent $i$ operates in accordance with this automaton. For clarity, elements of the reset map and brief descriptions of transitions are included.}
\label{fig:Automaton}
\end{figure} 
The coverage strategy for agent $i$ is represented by the hybrid automaton in Fig. \ref{fig:Automaton}. Rigorous definitions of all entities of the automaton, including the guard conditions and reset maps, are included in the appendix. Note that each agent operates in accordance with its own automaton and thus an arbitrary number of agents may be in any operating mode at any given time. Before proceeding, we provide a brief overview of each mode.\begin{itemize}\vspace{-3mm}
\item Local Coverage: This mode governs the active exploration of our surface of interest $\mathcal{C}$. When active, the agent continuously seeks to orient and translate $\mathcal{S}_i$ such that $\mathcal{S}_i$ intersects portions of $\mathcal{C}$ with a lower coverage level. This is conceptually similar to following the gradient of the coverage error. An agent currently assigned to an intruder may operate in Local Coverage while within an $\varepsilon_1$ bound of the the intruder's predicted impact point. Any agent not currently assigned to an intruder shall operate in Local Coverage assuming that it is within its assigned latitude partition. Operation in Local Coverage is always concurrent with agent assignment to the lowest concentric surface (see Surface Transfer below) and transition to Local Coverage can occur from any mode aside from Return to Base Mode.

\item Particle Intercept: In this mode, an intruder is assigned to agent $i$ and $i$ is guided along its assigned surface to the predicted impact point of the intruder. After intruder assignment occurs, the agent will nominally remain in Particle Intercept Mode until after the intruder impacts $\mathcal{C}$; however, the agent may temporarily leave the mode before impact to avoid collision through Surface Transfer Mode or to explore in Local Coverage within a $\varepsilon_1$ bound of the the intruder's predicted impact point.

\item Partition Transfer: This mode is defined for agents that are not currently assigned to an intruder and its purpose is to ensure that the agents are spatially distributed across the entire surface area of $\mathcal{C}$. Activation of this mode will guide agent $i$ along a longitudinally-oriented geodesic trajectory until its position satisfies a set of latitude constraints, i.e., agents travel to the southernmost latitude partition upon deployment and transition through progressively northern partitions as their fuel is depleted. Transition to this mode can occur from any other mode. The partitioning scheme is shown in Fig. \ref{fig:Partitions}.

\item Surface Transfer: This mode's primary purpose is to ensure that agents avoid collision with two distinct cases resulting in its activation. In the first case, two or more agents have violated a safe-proximity condition. The mode removes select agents from the deadlock by guiding them along vectors normal to $\mathcal{C}$ to a higher-altitude ellipsoidal surface concentric with $\mathcal{C}$. An agent trajectory is then temporarily confined to this newly assigned surface until it reaches the surface projection of its destination. The second case occurs under the condition that the agent has arrived at the projection of its destination on a higher-altitude surface. The mode is activated to return the agent to the innermost surface. Transition to this mode can occur from any other mode aside from Return to Base as agents in the latter mode always take priority in a deadlock. The surface transfer geometry is illustrated in Fig. \ref{fig:Surface Geometry}.

\item Return to Base: The final mode is activated when the time since an agent's deployment has surpassed some threshold. It guides the power-critical agent along the optimal trajectory to the refueling station. After charging, the agent is redeployed. Agent deployments occur one at a time with a fixed period. A power critical agent in Particle Intercept Mode or Surface Transfer Mode shall first complete its task of intercepting the assigned intruder or transferring surfaces before transitioning to Return to Base Mode. If an agent is designated as power-critical while in Partition Transfer Mode it shall immediately abandon its task and transition to Local Coverage which shall result in instantaneous transition to Return to Base Mode. Theoretical guarantees on successful return to base with respect to agent power lifespan in accordance with our automaton is presented in Theorem 2 of Section 5.
\end{itemize}\vspace{-3mm}
\section{Local Coverage Mode}\label{local coverage control}
\begin{figure}[h]
\centering
\includegraphics[width=.9\columnwidth,clip]{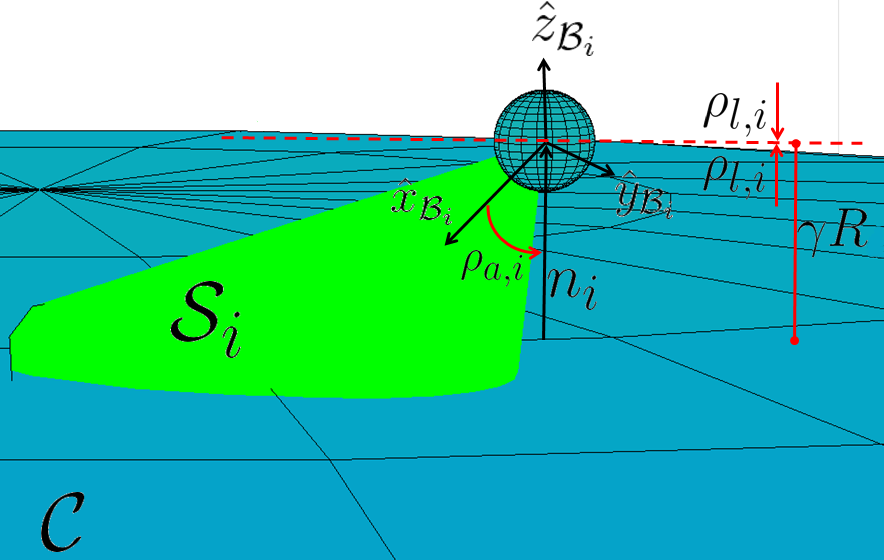}
\caption{As agent $i$ explores $\mathcal{C}$, $\rho_{l,i}$ is parallel to $n_i$ for $\lVert n_i \rVert<\gamma R$, antiparallel to $n_i$ for $\lVert n_i \rVert>\gamma R$, and the zero vector otherwise. This term prevents collision of $i$ with $\mathcal{C}$ and prevents $i$ from flying away from $\mathcal{C}$. $\rho_{a,i}$ tends to direct $\mathcal{S}_i$ onto $\mathcal{C}$.}
\label{fig:surface_explore}
\end{figure}

Local coverage constitutes the first of five hybrid modes in our automaton. This mode is gradient following in nature and commands agent $i$ to always seek to orient and translate $\mathcal{S}_i$ such that the volume of uncovered space intersecting $\mathcal{S}_i$ is increased. In this way, it emphasizes active exploration of the domain by agents that are not currently assigned to either monitor intruders or relocate within the domain. The control laws are designed such that agent motion in local coverage shall tend to reduce the rate of growth of the global coverage error. Define the global coverage error with respect to $C^{\star}$ as:
\vspace{-3mm}
\begin{align} \label{Global Coverage Error}
E(t)=\int\limits_{\mathcal{D}} \ h\left(C^{\star}C\left(\tilde{p}\right)-Q(t,\tilde{p})\right)\,d\tilde{p},
\end{align}
where $h(w)=(\max\{0,w\})^3$ with first derivative $h'=\frac{d h}{d w}=3(\max\{0,w\})^2$ and second derivative $h''=\frac{d^2 h}{d w^2}=6(\max\{0,w\})$. Our local coverage control laws are derived via differentiation of \eqref{Global Coverage Error}. This is included in the Appendix in the interest of space. The result is the selection of the following control strategy:
\small
\begin{subequations}
\label{coverage control law}
\begin{align}
u_i^{loc} &= k_{u}\frac{a_{i1}(t,Q(t,\tilde{p}))}{\sqrt{a_{i1}^2+a_{i2}^2+a_{i3}^2}}
+\hat{x}_{\mathcal{B}_i} \cdot \rho_{l,i} , \label{ui coverage} \\
v_i^{loc} &= k_{v}\frac{a_{i2}(t,Q(t,\tilde{p}))}{\sqrt{a_{i1}^2+a_{i2}^2+a_{i3}^2}}
+\hat{y}_{\mathcal{B}_i} \cdot \rho_{l,i}, \label{vi coverage} \\
w_i^{loc} &= k_{w}\frac{a_{i3}(t,Q(t,\tilde{p}))}{\sqrt{a_{i1}^2+a_{i2}^2+a_{i3}^2}}
+\hat{z}_{\mathcal{B}_i} \cdot \rho_{l,i}, \label{wi coverage} \\
r_i^{loc} &= \bar{r}_{i}sat\bigl(\frac{k_{r}a_{i4}(t,Q(t,\tilde{p}))}{\bar{r}_{i}}\bigr)+\hat{y}_{\mathcal{B}_i} \cdot \rho_{a,i}, \label{ri coverage} \\
s_i^{loc} &= \bar{s}_{i}sat\bigl(\frac{k_{s}a_{i5}(t,Q(t,\tilde{p}))}{\bar{s}_{i}}\bigr)+\hat{z}_{\mathcal{B}_i} \cdot \rho_{a,i}, \label{si coverage}
\end{align}
\end{subequations}
\normalsize
where: \vspace{-3mm}
\small
\begin{align}
\rho_{l,i}&=-\ln{\left(\frac{1}{\gamma R-\sr_i}\left(\lVert n_i \rVert-\sr_i\right)\right)}\mathcal{R}_1^{-1}\hat{n}_i,\\ 
\rho_{a,i}&=\xi \mathcal{R}_2^{-1} \begin{bmatrix}
0 \\
\arcsin{\left(\hat{n}_i \cdot \hat{z}_{\mathcal{G}}\right)}-\Theta_i \\
\atan2{\left(-\hat{n}_i \cdot \hat{y}_{\mathcal{G}},-\hat{n}_i \cdot \hat{x}_{\mathcal{G}}\right)}-\Psi_{i}\end{bmatrix}.
\end{align}
\normalsize
$\rho_{l,i}$ is a collision avoidance term with respect to the surface of interest. It takes a value of zero when agent $i$'s normalized distance from $\mathcal{C}$ is $\gamma R$ for $\gamma \in (0,1]$ and is logarithmically repulsive and attractive from the surface when the distance is decreased or increased respectively. Small values for $\gamma$ tend to direct the agent to travel closer to the surface. This coincides with a smaller cross section of $\mathcal{S}_i$ intersecting the surface but is also typically associated with a higher quality of sensing. A larger choice for $\gamma$ will direct the agent to fly at a higher altitude with respect to the surface and thus the area covered by $\mathcal{S}_i$ will tend to be broader with a decreased quality of sensing. $\rho_{a,i}$, for $\xi<<1$, encodes that the agents should tend to align $\hat{x}_{\mathcal{B}_i}$ with $-\hat{n}_i$ if the coverage terms associated with $r_i$ and $s_i$ have become sufficiently small. The physical meaning of $\rho_{a,i}$ is to direct $\mathcal{S}_i$ back onto $\mathcal{C}$ if it has reached a configuration in which it no longer intersects $\mathcal{C}$. See Fig. \ref{fig:surface_explore} for further details.

$\bar{r_i}$ and $\bar{s_i}$ are saturation limits for the coverage angular velocity inputs to the system. $k_u$, $k_v$ and $k_w$ are tuning gains which are chosen to satisfy $\sqrt{k_u^2+k_v^2+k_w^2} \leq U_{max}^{agt}$. As $\rho_{\ell,i}$ is normal to the surface, it can be shown that $U_{max}^{agt}$ is an upper bound to agent velocity tangential to $\mathcal{C}$.\vspace{-3mm}
\section{Particle Intercept Mode}\vspace{-3mm}
Assuming that particle $k$ is embedded within the surface upon impact, its position shall intersect $\mathcal{C}$ at most one time. We define particle $k$'s estimated impact time as \scriptsize$t_{ck}=\min{\left(t \in \mathbb{R}^{+} \mid \frac{\left(\hat{x}_k+\dot{\hat{x}}_kt\right)^2}{x_{\mathcal{C},r}^2} +\frac{\left(\hat{y}_k+\dot{\hat{y}}_kt\right)^2}{x_{\mathcal{C},r}^2}+\frac{\left(\hat{z}_k+\dot{\hat{z}}_kt\right)^2}{z_{\mathcal{C},r}^2}=1\right)}$\normalsize, with estimated impact point $\tilde{p}'_k\left(t_{ck}\right)=G\left(t_{ck}-t\right)\hat{\tilde{q}}_k\left(t\right)$. Upon detection, particle $k$ is assigned to a free agent $i$ with the minimum distance from the estimated point of impact along the $\hat{z}_\mathcal{G}$ direction. We define a new index, $i_k$, as the index of the agent assigned to intercept particle $k$ at destination $p_{id}=\tilde{p}'_k\left(t_{ck}\right)$: $i_k=\argmin_{i \in \{1,...,N\} \mid i_p \neq 1, f_i\neq 1}\lVert\bar{z}_k'(t_{ck})-z_i(t_{dk})\rVert$. Note that $\bar{z}_k'(t_{ck})$ is the $z$ component of $\bar{p}_k'(t_{ck})$ and $f_i \in \{0,1\}$ is a particle assignment flag for agent $i$ defined as $0$ when the agent is free (i.e., not currently assigned a particle). $i_p \in \{1,...,N\}$, the power index of agent $i$, shall be fully described in Section \ref{energy-aware domain partitioning}; however, it should be noted that the definition of $i_k$ implies that there are at most $N-1$ agents available for particle interception at any given time.\vspace{-2mm}

\begin{figure}[h]
\centering
\includegraphics[width=1\columnwidth,clip]{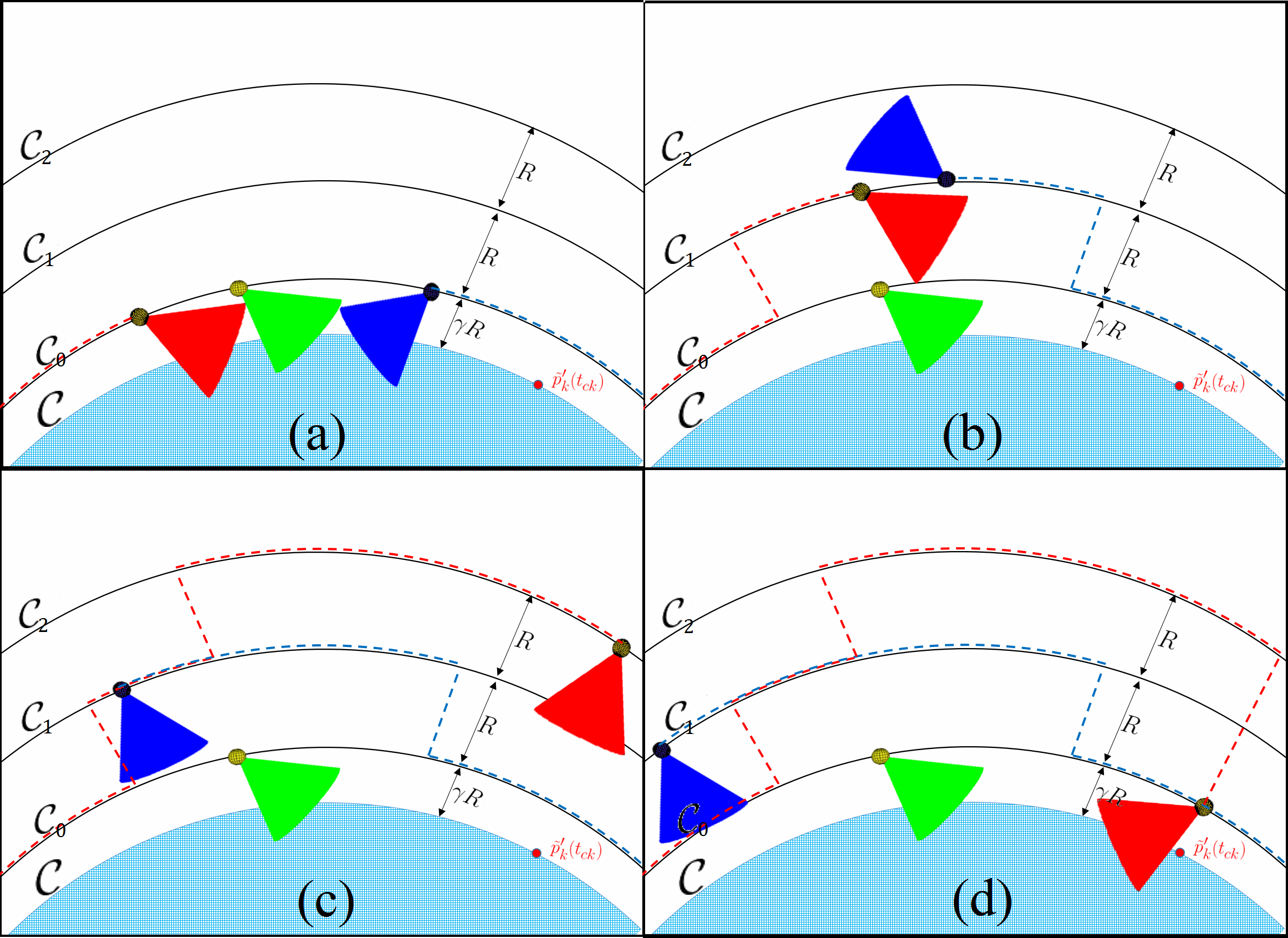}
\caption{Three agents enter a deadlock in (a). The green agent, which has the greatest time since deployment, is prioritized to continue on $\mathcal{C}_0$ and the red and blue agents are each transferred to $\mathcal{C}_1$ before entering a second deadlock in (b). The blue agent, which has the second greatest time since deployment, is prioritized to continue on $\mathcal{C}_1$ and the red agent is transferred to $\mathcal{C}_2$ before continuing along geodesic to $\vectorproj[\mathcal{C}_2]{\tilde{p}_k'(t_{ck})}$ in (c). Red agent transfers back to $\mathcal{C}_0$ directly above predicted impact point of particle $k$ in (d). Note that surface transfer trajectories are always normal to $\mathcal{C}_{\mu_i}, \, \forall \mu_i \in \{0,...,N-1\}$.}
\label{fig:Surface Geometry}
\end{figure}

As with our local coverage strategy, it is assumed that agents shall maintain a distance $\gamma R$ normal to $\mathcal{C}$ in the nominal case that they are not maneuvering to avoid collision. We define an ellipsoid of revolution, $\mathcal{C}_{0}$, which is concentric with $\mathcal{C}$ and has the property that each semi-principal axis is $\gamma R$ longer than its associated counterpart in $\mathcal{C}$, i.e., $x_{\mathcal{C}_0,r}=x_{\mathcal{C},r}+\gamma R$, and $z_{\mathcal{C}_0,r}=z_{\mathcal{C},r}+\gamma R$. The nominal trajectories of $i$ are attractive to $\mathcal{C}_0$. Agents maneuvering to avoid collision shall transfer to additional concentric ellipsoidal surfaces each separated by a distance of $R$. These surfaces are denoted $\mathcal{C}_1, \mathcal{C}_2,...,\mathcal{C}_{N-1}$ with associated semi-principal axes $x_{\mathcal{C}_1,r}=x_{\mathcal{C},r}+\left(\gamma+1\right) R$ and $z_{\mathcal{C}_1,r}=z_{\mathcal{C},r}+\left(\gamma+1\right) R$, $x_{\mathcal{C}_2,r}=x_{\mathcal{C},r}+\left(\gamma+2\right) R$ and $z_{\mathcal{C}_2,r}=z_{\mathcal{C},r}+\left(\gamma+2\right) R$, etc. Surface assignment and transfer scheduling in collision avoidance mode is described in full detail in Section \ref{Surface Transfer Mode} and the geometry is illustrated in Fig. \ref{fig:Surface Geometry}.

When agent $i$ has been assigned to intercept particle $k$, $f_i$ is set to $1$ and it is said to have transitioned into particle intercept mode. In this mode, agent $i$ shall nominally follow the optimal trajectory along $\mathcal{C}_0$ to within a $\varepsilon_1$ bound of the projection of point $\tilde{p}_k'(t_{ck})$ onto $\mathcal{C}_0$ (denoted $\vectorproj[\mathcal{C}_0]{\tilde{p}_k'(t_{ck})}$). The agent shall then transition to local coverage to actively explore within this $\varepsilon_1$ bound until $t>t_{ck}$ at which time $f_i$ is set to $0$. If local coverage guides the agent out of the $\varepsilon_1$ bound, particle intercept mode will again guide the agent back inside the bound. The optimal trajectory is referred to as a geodesic and its computation may be executed in an iterative manner. Specifically, we use Vincenty's formulae as presented in \citet{Vincenty_1975}. For cases involving nearly antipodal points in which the standard inverse method does not converge, we use Vincenty's supplemental algorithm presented in \citet{Vincenty_Supplement}.\vspace{-2mm}

\begin{figure}[h]
\centering
\includegraphics[width=.7\columnwidth,clip]{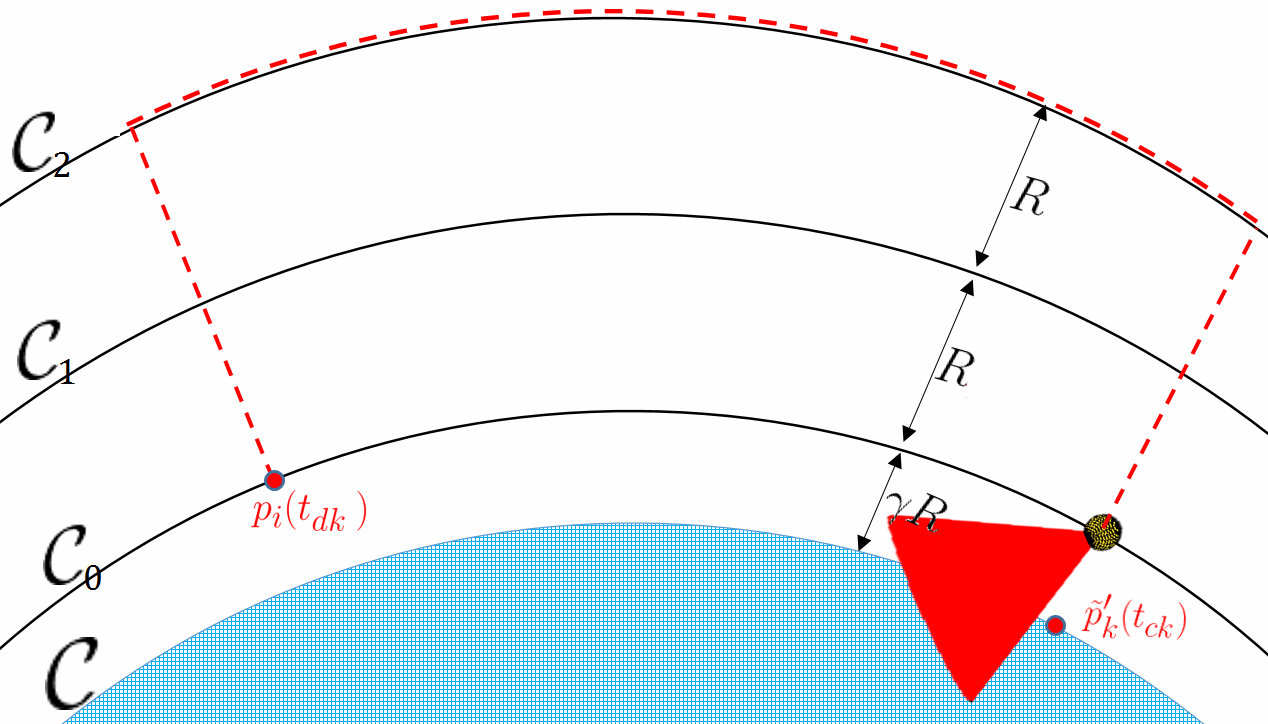}
\caption{The longest possible path from $p_i(t_{dk})$ to $\tilde{p}_k'(t_{ck})$, taken by agent $i$ assigned to intercept particle $k$, is illustrated above. We denote this path as $\mathcal{P}_{max}$ and it may be upper bounded as established in Lemma 1.}
\label{fig:Proof_Fig}
\end{figure}

As an input, Vincenty's algorithm requires an ellipsoid of revolution along with two points, current and desired position, on that surface. The algorithm returns a heading angle measured clockwise from North. This heading angle shall be referred to as $\chi_i$. We now define the heading unit vector $\hat{\nu}_i$ which lies in a plane tangent to the surface at $p_i$. It may be computed by rotating the North-pointing vector at $p_i$ clockwise by an angle of $\chi_i$ within the tangent plane. For our implementation of Vincenty's algorithm, we input the following: $\mathcal{C}_{\mu_i}$ for surface assignment index $\mu_i \in \{0,...,N-1\}$, $p_i$, and $\vectorproj[\mathcal{C}_{\mu_i}]{\tilde{p}_k'(t_{ck})}$.
The position controller used to guide agent $i$ to $\vectorproj[\mathcal{C}_{\mu_i}]{\tilde{p}_k'(t_{ck})}$ is composed of two terms: one which commands velocity tangential to $\mathcal{C}_{\mu_i}$ along $\hat{\nu}_i$ and one logarithmic term which commands velocity normal to $\mathcal{C}_{\mu_i}$ in order to constrain the geodesic trajectory of $i$ to $\mathcal{C}_{\mu_i}$. The particle intercept mode position control law is:\vspace{-6mm}
\small
\begin{align}
\label{particle intercept position control}
 \begin{bmatrix}
 u_{i}^{pim} \\
 v_{i}^{pim} \\
 w_{i}^{pim} \\
 \end{bmatrix}
 =U_{max}^{agt}\mathcal{R}_{1}^{-1}\frac{\hat{\nu}_i-\ln{\left(\frac{1}{\left(\gamma+\mu_i\right) R-\sr_i}\left(\lVert n_i \rVert-\sr_i\right)\right)}\hat{n}_i}{\lVert \hat{\nu}_i-\ln{\left(\frac{1}{\left(\gamma+\mu_i\right) R-\sr_i}\left(\lVert n_i \rVert-\sr_i\right)\right)}\hat{n}_i\rVert}.
\end{align}
\normalsize
As agent $i$ travels along the geodesic, it is desirable that it should point $\mathcal{S}_i$ towards $\mathcal{C}$. Therefore, the orientation controller for particle intercept mode is similar to that of Section \ref{local coverage control}: \vspace{-3mm}
\begin{align}
\label{particle intercept orientation control}
 \begin{bmatrix}
 q_{i}^{pim} \\
 r_{i}^{pim} \\
 s_{i}^{pim} \\
 \end{bmatrix}
 = 
\mathcal{R}_2^{-1} \begin{bmatrix}
0 \\
\arcsin{\left(\hat{n}_i \cdot \hat{z}_{\mathcal{G}}\right)}-\Theta_i \\
\atan2{\left(-\hat{n}_i \cdot \hat{y}_{\mathcal{G}},-\hat{n}_i \cdot \hat{x}_{\mathcal{G}}\right)}-\Psi_{i}\end{bmatrix},  
\end{align}
which is essentially a proportional controller that tends to align $\hat{x}_{\mathcal{B}_i}$ with $-\hat{n}_i$.
As \eqref{particle intercept position control} commands the vehicle to follow the optimal length path along $\mathcal{C}_{\mu_i}$ to $\vectorproj[\mathcal{C}_{\mu_i}]{\tilde{p}_k'(t_{ck})}$, we can establish a few guarantees on system performance. To simplify notation, define: \vspace{-3mm}
\begin{align}
\label{series}
g_{\mathcal{C}_{N-1}}=\left[1+
\sum_{n=1}^{\infty} \left(\frac{\left(2n-1\right)!!}{2^nn!}\right)^2 \frac{\left(\frac{x_{\mathcal{C}_{N-1},r}-z_{\mathcal{C}_{N-1},r}}{x_{\mathcal{C}_{N-1},r}+z_{\mathcal{C}_{N-1},r}}\right)^{2n}}{\left(2n-1\right)^2} 
\right],
\end{align} and $g_{\mathcal{C}_{0}}$ is defined similarly in terms of the semi-principal axes of $\mathcal{C}_0$.
\begin{Lemma1}
\label{Lemma1}
\textnormal{Let us assume that agent $i$ has been assigned to particle $k$ with $f_i:=1$. Given an arbitrary agent position $p_i(t_{dk})$ and an arbitrary predicted impact point for the intruder $\tilde{p}_k'(t_{ck})$, there exists an upper bound to the maximum path length until interception: $\mathcal{P}_{max} \leq \pi x_{\mathcal{C}_{N-1},r}+\frac{\pi}{2}\left(x_{\mathcal{C}_{N-1},r}+z_{\mathcal{C}_{N-1},r}\right)g_{\mathcal{C}_{N-1}}+2\left(N-1\right)R$.}
\begin{pf}
\textnormal{At the moment that $f_i:=1$ we have that agent $i$ transitions to Particle Intercept Mode. Under the condition that the agent has not yet come within proximity of the predicted impact point, i.e., $\lVert p_i-\vectorproj[\mathcal{C}_{\mu_i}]{\tilde{p}_k'(t_{ck})}  \rVert > \varepsilon_1$, we have that only $G\left(\zeta_{i2},\zeta_{i4}\right)$ and $G\left(\zeta_{i4},\zeta_{i2}\right)$ are defined (see Appendix). These two transitions occur sequentially for each deadlock event that agent $i$ encounters as it approaches $\vectorproj[\mathcal{C}_{\mu_i}]{\tilde{p}_k'(t_{ck})}$.}\vspace{-2mm}

\textnormal{In any given deadlock arrangement, one agent remains on its current surface without ascending to a higher one. This implies that $\mu_i=1$ for at most $N-1$ agents as the remaining agent would be unable to encounter a deadlock on $\mathcal{C}_0$. Furthermore, this implies that $\mu_i=2$ for at most $N-2$ agents etc. until we have $\mu_i=N$ for at most zero agents. The worst cast surface assignment that can be incurred during sequential cycles of $\left(\left(\zeta_{i2},\zeta_{i4}\right),\left(\zeta_{i4},{i2}\right)\right)$ would be $\mu_i=N-1$.}\vspace{-2mm}

\textnormal{As the geodesic path length between any two points projected onto surface $\mathcal{C}_{\mu_i}$ shall always be less than the geodesic path length between the same two points projected onto surface $\mathcal{C}_{\mu_i+1}$, we may bound the geodesic portion of the trajectory by one that is constrained entirely to $\mathcal{C}_{N-1}$. We denote this term $\mathcal{P}_{geo}$. As any two points on $\mathcal{C}_{\mu_i}$ can be connected by a path of constant latitude $\mathcal{P}_{lat}$ followed by a path of constant longitude $\mathcal{P}_{long}$, we have that:}\vspace{-3mm}
\begin{align}
\label{path upper bound}
\mathcal{P}_{geo} \leq \mathcal{P}_{lat}+\mathcal{P}_{long}.
\end{align}
\textnormal{For two generic points on $\mathcal{C}_{N-1}$, we have that:}\vspace{-3mm}
\begin{align}
\label{lat upper bound}
\mathcal{P}_{lat} &\leq \pi x_{\mathcal{C}_{N-1},r}, \\
\label{long upper bound}
\mathcal{P}_{long} &\leq \frac{\pi}{2}\left(x_{\mathcal{C}_{N-1},r}+z_{\mathcal{C}_{N-1},r}\right)g_{\mathcal{C}_{N-1}}.
\end{align}
\textnormal{where the bound on $\mathcal{P}_{lat}$ is half of the circumference of the ellipsoid of revolution $\mathcal{C}_{N-1}$ about its equator and the bound on $\mathcal{P}_{long}$ is half of the perimeter of the revolved ellipse. The infinite series expression term, denoted $g_{\mathcal{C}_{N-1}}$ in \eqref{long upper bound}, is first presented in \citet{ivory_1798}. The remaining portion of the path length is simply the straight line segments connecting $\mathcal{C}_0$ to $\mathcal{C}_{N-1}$ and back again. This length is precisely $2\left(N-1\right)R$. Thus, $\mathcal{P}_{max}=\mathcal{P}_{geo}+2\left(N-1\right)R$ as illustrated in Fig. \ref{fig:Proof_Fig}. Invoking \eqref{lat upper bound} and \eqref{long upper bound} gives us $\mathcal{P}_{max} \leq \pi x_{\mathcal{C}_{N-1},r}+\frac{\pi}{2}\left(x_{\mathcal{C}_{N-1},r}+z_{\mathcal{C}_{N-1},r}\right)g_{\mathcal{C}_{N-1}}+2\left(N-1\right)R$. This concludes the proof.} \vspace{-2mm}
\end{pf}
\end{Lemma1}
 \vspace{-2mm}
\begin{Theorem_interception}
\label{Theorem_interception}
\textnormal{Assuming that the bounds on intruder velocity and range from detection to impact satisfy $\frac{R_{det}}{U_{max}^{int}} > \frac{\mathcal{P}_{max}}{U_{max}^{agt}}$, where $\mathcal{P}_{max}$ may be bounded via Lemma 1, agent $i$ shall reach $\vectorproj[\mathcal{C}_{0}]{\tilde{p}_k'(t_{ck})}$ before $t_{ck}$.} \vspace{-2mm}
\begin{pf}
\textnormal{Given the fact that agents in the particle intercept and surface transfer modes travel at speed $U_{max}^{agt}$, we have that the time $t_{ik}$ required to travel from $p_i\left(t_{dk}\right)$ to $\vectorproj[\mathcal{C}_{0}]{\tilde{p}_k'(t_{ck})}$ must satisfy: $t_{ik} \leq \frac{\mathcal{P}_{max}}{U_{max}^{agt}}$. Given that $t_{ck}-t_{dk} \geq \frac{R_{det}}{U_{max}^{int}}$, agent $i$ reaching $\vectorproj[\mathcal{C}_{0}]{\tilde{p}_k'(t_{ck})}$ before $t_{ck}$ implies that $\frac{R_{det}}{U_{max}^{int}}>t_{ik}$. This is guaranteed if $\frac{R_{det}}{U_{max}^{int}} > \frac{\mathcal{P}_{max}}{U_{max}^{agt}}$  This concludes the proof.}
\end{pf}
\end{Theorem_interception}
\begin{Remark1}
\label{Remark1}
\textnormal{At any given time, there are at most $N-1$ agents available to intercept particles. Thus, satisfaction of Theorem \ref{Theorem_interception} implies that the network is capable of intercepting all particles so long as a maximum of $N-1$ particle impacts occur in any moving time window of $\frac{\pi x_{\mathcal{C}_{N-1},r}+ \frac{\pi}{2}\left(x_{\mathcal{C}_{N-1},r}+z_{\mathcal{C}_{N-1},r}\right)g_{\mathcal{C}_{N-1}}+2\left(N-1\right)R}{U_{max}^{agt}}$.}
\end{Remark1}\vspace{-3mm}
\section{Energy-aware Scheduling Protocol}\vspace{-3mm}
\label{energy-aware domain partitioning}
\subsection{Domain Partitioning}\vspace{-3mm}
As this is a persistent coverage protocol, which operates indefinitely, it is necessary to establish an agent deployment and scheduling protocol that realistically considers the agents' finite power and/or propulsive resources. Our strategy is to periodically deploy agents from a fueling station $\mathcal{F}$ which we assume to be located at the North pole of $\mathcal{C}_0$, i.e., at the point $\left[0\;\;0\;\;z_{\mathcal{C}_0,r}\right]^T$. Define $T^{\star}$ as the power lifespan of each agent in the network. Given $T^{\star}$ and $N$, we define our deployment and scheduling protocol such that one agent is deployed from $\mathcal{F}$ every $\frac{T^{\star}}{N}$ seconds. The initial deployment is that of agent $i=1$ at $t=0$ seconds with agent $i=2$ following at $t=\frac{T^{\star}}{N}$. This continues indefinitely with the second deployment of agent $i=1$ occurring at $t=T^{\star}$ seconds.\vspace{-2mm}

In order to adequately distribute agents across $\mathcal{C}$, it is desirable to partition the domain and assign agents to monitor separate regions. Specifically, partitioning the domain by latitude, rather than longitude, ensures that agents are poised to intercept particles without the need for frequent crossings of the equator which tend to be associated with larger values of $\mathcal{P}_{geo}$ on an oblate spheroid.\vspace{-2mm}

Define the power index of agent $i$ as $i_p(t)=1+\mod{\left(i-2-\left\lfloor\frac{tN}{T^{\star}}\right\rfloor,N\right)}$ where the first argument of our modulo operation is the dividend and the second argument is the divisor. The lower-bracketed delimiters represent the floored division operation. Upon deployment from $\mathcal{F}$, agent $i$ has power index $i_p=N$ and this index is reduced by one every $\frac{T^{\star}}{N}$ seconds until $i_p=1$, i.e., agent $i$ is the power critical agent. Note that no two agents may share the same power index as a result of our periodic deployment and scheduling protocol.\vspace{-2mm}

\begin{figure}[h]
\centering
\includegraphics[width=.9\columnwidth,clip]{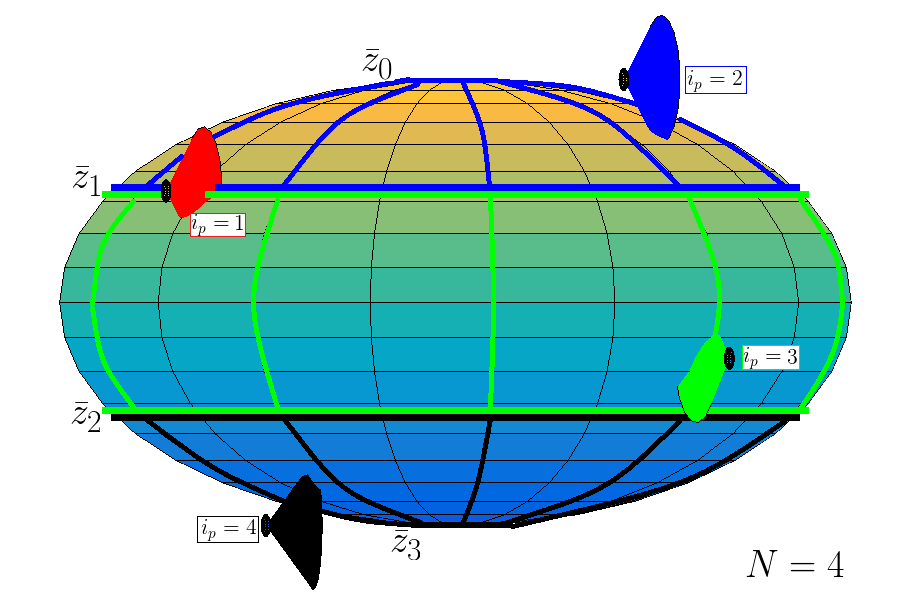}
\caption{Our domain partitioning scheme for $\mathcal{C}$ is illustrated above. Agents with $i_p \in \{2,3,4\}$ are indicated with blue, green and black $\mathcal{S}_i$ respectively. Their partitions are separated by latitude lines upper bounded at $\bar{z}_{{i_p}-2}$ and lower bounded at $\bar{z}_{{i_p}-1}$. The power critical has $\mathcal{S}_i$ indicated in red.}
\label{fig:Partitions}
\end{figure}

Latitude partitions are characterized by a static upper bound in $\hat{z}_{\mathcal{G}}$ denoted $\bar{z}_{i_p-2}$ and a static lower bound $\bar{z}_{i_p-1}$. Rather than dynamically sizing partitions relative to agent power resources, we divide partitions such that $N-1$ agents are assigned equal surface areas of $\mathcal{C}$ to explore. This choice maximizes the coverage of any individual partition as the allocation of a larger partition to a recently deployed agent would result in less effective coverage of that partition. Agents are sorted by their remaining power and transfer between partitions that are progressively closer to $\mathcal{F}$ as their power resource expires. Define the surface area of our ellipsoid of revolution $\mathcal{C}$ as:\vspace{-10mm}
\small
\begin{align}
\label{surface area C}
A_{\mathcal{C}}=2\pi x_{\mathcal{C},r}^2\left(1+\frac{1+\left(1-\frac{z_{\mathcal{C},r}^2}{x_{\mathcal{C},r}^2}\right)}{\left(\sqrt{1-\frac{z_{\mathcal{C},r}^2}{x_{\mathcal{C},r}^2}}\right)}\artanh{\left(\sqrt{1-\frac{z_{\mathcal{C},r}^2}{x_{\mathcal{C},r}^2}}\right)}\right).
\end{align}
\normalsize
The agent with $i_p=2$ is assigned to monitor the partition characterized by upper bound at north pole of $\mathcal{C}$, i.e., $\bar{z}_{0}=z_{\mathcal{C},r}$. The lower bound $\bar{z}_1$ may be computed by dividing \eqref{surface area C} by $\left(N-1\right)$, equating with the integral of ellipse cross sectional circumferences parametrized by $\tilde{z}$, and then numerically solving for $\bar{z}_1$:\vspace{-3mm}
\tiny
\begin{align}
\label{Partition 1 Bound}
\frac{A_{\mathcal{C}}}{N-1}=\int_{z_{\mathcal{C},r}}^{\bar{z}_1}
2 \pi \sqrt{\left(x_{\mathcal{C},r}^2-\frac{x_{\mathcal{C},r}^2\tilde{z}^2}{z_{\mathcal{C},r}^2}\right)\left(1+\frac{\tilde{z}^2x_{\mathcal{C},r}^4}{x_{\mathcal{C},r}^2\left(z_{\mathcal{C},r}^4-z_{\mathcal{C},r}^2\tilde{z}^2\right)}\right)}d\tilde{z}.
\end{align} 
\normalsize
One may then iteratively solve for the remaining bounds for increasing values of $i_p$ up to $i_p=N-1$:\vspace{-3mm}
\tiny
\begin{align}
\label{Generic Partition Bound}
\frac{A_{\mathcal{C}}}{N-1}=\int_{\bar{z}_{i_p-2}}^{\bar{z}_{i_p-1}}
2 \pi \sqrt{\left(x_{\mathcal{C},r}^2-\frac{x_{\mathcal{C},r}^2\tilde{z}^2}{z_{\mathcal{C},r}^2}\right)\left(1+\frac{\tilde{z}^2x_{\mathcal{C},r}^4}{x_{\mathcal{C},r}^2\left(z_{\mathcal{C},r}^4-z_{\mathcal{C},r}^2\tilde{z}^2\right)}\right)}d\tilde{z}.
\end{align}
\normalsize
The final computation of \eqref{Generic Partition Bound} for $i_p=N$ is not necessary as $\bar{z}_{N-1}$ is the south pole of $\mathcal{C}$, i.e., $\bar{z}_{N-1}=-z_{\mathcal{C},r}$, although this may be shown through numerical computation as well. Our partitioning strategy for the case where $N=4$ is presented in Fig. \ref{fig:Partitions}.

Note that no partition has been assigned to the agent for which $i_p=1$. This is the power critical agent and it shall have flag $f_i:=1$ at the instant $i_p:=1$. The power critical agent cannot be assigned a new particle to intercept after $i_p:=1$ as this opens the possibility that particle assignment could occur near the end of the $\frac{T^{\star}}{N}$ time window during which time  the agent with $i_p=1$ should be transitioning back to $\mathcal{F}$ to exchange its power source. The power critical agent will instead spend the majority of this time window in local coverage mode assisting the other agents in gathering information. It can only be tasked with intercepting a particle if this assignment had occurred previously when $i_p=2$. In this scenario, the agent should be capable of intercepting particle $k$ and then transitioning back to $\mathcal{F}$ so long as a bound is established on the length of our deployment scheduling window $\frac{T^{\star}}{N}$.
\begin{Theorem_window}
\label{Theorem_window}
\textnormal{If agent power lifespan $T^{\star}$ satisfies $\frac{T^{\star}}{N}\geq t_{ck}-t_{dk}+\frac{\pi}{2U_{max}^{agt}}\left(x_{\mathcal{C}_{0},r}+z_{\mathcal{C}_{0},r}\right)g_{\mathcal{C}_{0}}, \; \forall k$ then the agent with $i_p=1$ shall always be capable of reaching $\mathcal{F}$ within $\frac{T^{\star}}{N}$ of the time at which $i_p:=1$.}

\begin{pf}
\textnormal{Consider the worst-case scenario in which the agent with $i_p=2$ is assigned to intercept particle $k$ at the instant before $i_p:=1$. It's remaining flight time is currently $\frac{T^{\star}}{N}$. The time required to intercept the particle is $t_{ck}-t_{dk}$, after which our control strategy dictates that the agent will follow a geodesic trajectory to $\mathcal{F}$. As $\mathcal{F}$ lies at the north pole of $\mathcal{C}_0$, this will be a trajectory of constant longitude which may be upper bounded by a length half the perimeter of our revolved ellipsoid: $\frac{\pi}{2}\left(x_{\mathcal{C}_0,r}+z_{\mathcal{C}_0,r}\right)g_{\mathcal{C}_0}$ by definition. As the agent is controlled by \eqref{particle intercept position control} with a North-pointing $\hat{\nu}_i$, it will proceed along this geodesic at speed $U_{max}^{agt}$. Thus the time required to complete this trajectory is $\frac{\pi}{2U_{max}^{agt}}\left(x_{\mathcal{C}_0,r}+z_{\mathcal{C}_0,r}\right)g_{\mathcal{C}_0}$ and we may bound our deployment window: $\frac{T^{\star}}{N}\geq t_{ck}-t_{dk}+\frac{\pi}{2U_{max}^{agt}}\left(x_{\mathcal{C}_0,r}+z_{\mathcal{C}_0,r}\right)g_{\mathcal{C}_0}, \; \forall k$. This concludes the proof.}
\end{pf}
\end{Theorem_window}
\begin{Remark2}
\label{Remark2}
\textnormal{The appropriate design method for this surveillance system is to first ensure that the time from detection to impact of any arbitrary particle, $t_{ck}-t_{dk}$, as governed by the omnidirectional range sensor satisfies Theorem \ref{Theorem_interception}. One must subsequently ensure that power lifespan $T^{\star}$, for all agents, satisfies Theorem \ref{Theorem_window}.}
\end{Remark2}\vspace{-3mm}
\subsection{Partition Transfer and Return to Base} \vspace{-3mm}
If an agent with $i_p \in \{2,...,N\}$ lies outside of its prescribed partition, and we have $i_f=0$, then the agent shall enter partition transfer mode. This mode uses the same geodesic position and orientation controllers \eqref{particle intercept position control} and \eqref{particle intercept orientation control} with the destination position set to the point: \vspace{-5mm}
\footnotesize
\begin{align*}
&p_{id}=\left[x_{id} \; \; y_{id} \; \; z_{id}\right]^T= \\
&\begin{bmatrix}
x_{\mathcal{C},r}\cos\left(\arcsin\left(\frac{z_{id}}{z_{\mathcal{C},r}}\right)\right)\cos\biggl(\atan2\bigl(y_i(t),x_i(t)   \bigr)\biggr)\\
y_{\mathcal{C},r}\cos\left(\arcsin\left(\frac{z_{id}}{z_{\mathcal{C},r}}\right)\right)\sin\biggl(\atan2\bigl(y_i(t),x_i(t)   \bigr)\biggr)\\
\bar{z}_{i_p-1}, \text{if $z_i<\bar{z}_{i_p-1}$}; \text{or } \bar{z}_{i_p-2}, \text{if $z_i>\bar{z}_{i_p-2}$} 
\end{bmatrix},  
\end{align*}
\normalsize
i.e., the closest point along the agent's current longitude which lies on the boundary of its assigned partition. 

The return to base mode is similar to partition transfer mode but is defined for the agent with $i_p=1$. This mode is activated when the time since agent $i$'s last deployment from $\mathcal{F}$, denoted $t_{i\mathcal{F}} \geq T^{\star}-\frac{\pi}{2U_{max}^{agt}}\left(x_{\mathcal{C}_0,r}+z_{\mathcal{C}_0,r}\right)g_{\mathcal{C}_0}$ as established in Theorem \ref{Theorem_window}. The control strategy is the same as partition transfer mode with the desired position set to $\mathcal{F}$. Control laws for partition transfer mode and return to base shall be denoted with superscripts $ptm$ and $rtb$ respectively. \vspace{-3mm}
\section{Surface Transfer Mode}\vspace{-3mm}
\label{Surface Transfer Mode}
The primary purpose of surface transfer mode is to encode collision avoidance and it can be transitioned into from any other mode aside from the return to base mode. This mode is triggered for agent $i$, assigned to surface $\mathcal{C}_{\mu_i}$, when we have the condition that $\lVert p_i-p_j \rVert \leq R$ for $i \neq j$. Denote $\tilde{j}=i \cup j$ as the set of agents satisfying this condition. Agents in $\tilde{j}$ are ranked by $t_{\tilde{j}\mathcal{F}}$. One agent, denoted $i_{pr}$, whose value for $t_{\tilde{j}\mathcal{F}}$ is highest, i.e., $i_{pr}=\argmax_{\tilde{j}}\left(t_{\tilde{j}\mathcal{F}}\right)$ is permitted to proceed. The remaining agents increment their surface assignment indices, $\mu_i$, by one and transition to surface transfer mode. This mode controls the agents to follow $\hat{n}_i$ until they have transferred to their newly assigned concentric surface at a height $R$ above the previous. Note that in general, convexity of surface $\mathcal{C}$ is required to ensure that intersections of $n_i$ and $n_j, \, \forall i \neq j,$ lie within the interior space that is bounded by the surface. The surface transfer position control strategy is: \vspace{-3mm}
\begin{align}
\label{avoidance position control}
 \begin{bmatrix}
 u_{i}^{stm} \\
 v_{i}^{stm} \\
 w_{i}^{stm} \\
 \end{bmatrix}
 =U_{max}^{agt}\mathcal{R}_{1}^{-1}\frac{\ln{\left(\frac{1}{\left(\gamma+\mu_i\right) R-\sr_i}\left(\lVert n_i \rVert-\sr_i\right)\right)}\hat{n}_i}{\lVert \ln{\left(\frac{1}{\left(\gamma+\mu_i\right) R-\sr_i}\left(\lVert n_i \rVert-\sr_i\right)\right)}\hat{n}_i\rVert}.
\end{align}
As the agents ascend to a point at which $R$ does not intersect $\mathcal{C}$, sensing information is not gathered in avoidance mode and thus the avoidance orientation control is simply $\left[q_i^{stm}\;\;r_i^{stm}\;\;s_i^{stm}\right]^{T}=\left[0\;\;0\;\;0\right]^T$. 

Agents are said to have converged upon their newly assigned surface when $\lvert\ln{\frac{\lVert n_i \rVert -\sr_i}{\left(\gamma+\mu_i\right)R-\sr_i}}\rvert<\varepsilon_2$. At this point, each agent shall transition back to its prior mode as described in the following two scenarios.
\begin{enumerate}
\item If agent $i$ had been in either particle intercept or partition transfer mode before the deadlock, it shall resume that mode and continue along a geodesic trajectory on the newly assigned surface until it reaches the projection of its destination. At this point, the condition that $\lVert p_i-\vectorproj[\mathcal{C}_{\mu_i}]{p_{id}} \rVert \leq \varepsilon_1$ triggers a reset $\mu_i:=0$ concurrent with a transition back to surface transfer mode thus allowing the agent to transfer back to $\mathcal{C}_0$. The agent then resumes coverage of $\mathcal{C}_0$ in its prior mode. For additional details on flag conditions in these transitions, see guards $G\left(\zeta_{i2},\zeta_{i4}\right), G\left(\zeta_{i4},\zeta_{i2}\right), G\left(\zeta_{i3},\zeta_{i4}\right)$, and $G\left(\zeta_{i4},\zeta_{i3}\right)$ of our hybrid automaton as presented in the appendix. 

\item If agent $i$ had been in local coverage mode before the deadlock, it shall then transition back to local coverage mode concurrent with reset $\mu_i:=0$. This transition is dependent upon the conditions that $f_i=0$ and that the agent is operating within its assigned partition. The agent shall oscillate between local coverage and surface transfer at an altitude of $R$ above $\mathcal{C}_0$ until $i_{pr}$ has moved along $\mathcal{C}_0$ to resolve the deadlock. At this point, the local coverage controller shall attract agent $i$ back to the surface.
\end{enumerate}
While similar work on multi-agent systems often invoke avoidance barrier functions to encode collision avoidance, such as in \citet{Global_Avoidance}, it may be impossible to bound the time that agents spend avoiding one another in these maneuvers\textemdash especially when the algorithm
is scaled to many agents. In contrast, our technique results in an explicit bound on path length to an intruder as was proven in Lemma \ref{Lemma1}. With an additional assumption on the size of agents, we can establish a guarantee on collision avoidance for agents in surface transfer mode.\vspace{-2mm}
\begin{Theorem_avoidance}
\label{Theorem_avoidance}
\textnormal{For agents $\{i,j\} \in \tilde{j}$, the condition that $\min(R_{\tilde{j}})>2\sr_{i}+2\sr_{j}$ implies that $i$ does not collide with $j$.}\vspace{-2mm}
\begin{pf}
\textnormal{Consider the case in which $i\neq i_{pr}$ and $j\neq i_{pr}$. Both agents operate in accordance with \eqref{avoidance position control} and follow trajectories along $\hat{n}_i$ and $\hat{n}_j$ respectively. Both unit vectors are normal to surface $\mathcal{C}_{\mu_i}$, an ellipsoid of revolution, and thus diverge from one another away from $\mathcal{C}_{\mu_i}$. Agents $i$ and $j$ shall enter surface transfer mode at an instant when $\lVert p_i-p_j \rVert \geq \min\left(R_{\tilde{j}}\right)$ and their distance shall tend to increase under \eqref{avoidance position control}. Thus $\min(R_{\tilde{j}})>\sr_{i}+\sr_{j}$ and subsequently $\min(R_{\tilde{j}})>2\sr_{i}+2\sr_{j}$ imply that they avoid collision.}\vspace{-2mm}

\textnormal{Consider the case in which $i=i_{pr}$ and thus $j\neq i_{pr}$. In the instant that $j$ transitions to surface transfer mode we have that $\lVert p_i-p_j \rVert\geq \min\left(R_{\tilde{j}}\right)$. Thus the distance for $i$ to travel until collision is greater than or equal to $\min\left(R_{\tilde{j}}\right)-\sr_i-\sr_j$. This straight line path for $i$ is a conservative estimate as the true path is curved. Collision will be avoided if agent $j$, whose path is normal to the surface, may cover a distance $\sr_i+\sr_j$ before $i$ covers $\min\left(R_{\tilde{j}}\right)-\sr_i-\sr_j$. As $j$ moves at speed $U_{max}^{agt}$ and $i$'s tangential speed is upper bounded by $U_{max}^{agt}$, this condition is satisfied if
$\min\left(R_{\tilde{j}}\right)-\sr_i-\sr_j > \sr_i+\sr_j$. This may equivalently be written as $\min\left(R_{\tilde{j}}\right) >2\sr_i+2\sr_j$. These arguments apply to the case in which $j=i_{pr}$ and $i\neq i_{pr}$ as well. This concludes the proof.}\vspace{-2mm}
\end{pf}
\end{Theorem_avoidance}\vspace{-5mm}
\section{Simulations}\vspace{-3mm}
\begin{figure*}
\centering
\includegraphics[width=\textwidth]{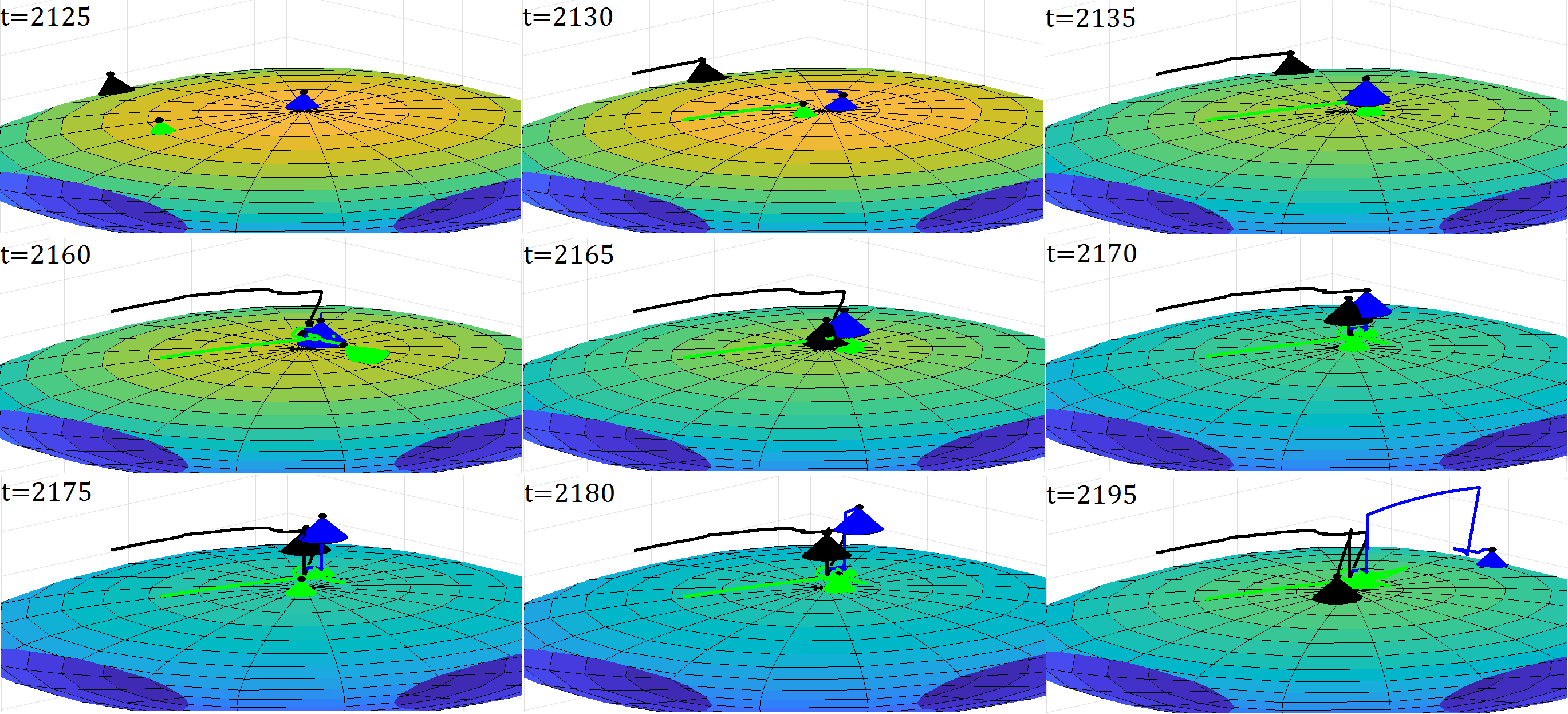}
\caption{Agent $i=3$, indicated with green $\mathcal{S}_i$, is on a collision course with agent $i=1$, indicated with blue $\mathcal{S}_i$, during the interval from $t=2125$ to $t=2130$. At $t=2135$, agent $i=1$ has transitioned to surface transfer mode and is following a trajectory normal to the surface while agent $i=4$, indicated with black $\mathcal{S}_i$, follows a collision course through $t=2160$. Agent $i=4$ transitions to surface transfer mode as well leading to the conditions that $\mu_1=2$ and $\mu_4=1$, i.e., agent $i=1$ is assigned to the second tier of avoidance surfaces at a higher altitude than $i=4$ as illustrated at $t=2180$. Both agents proceed along their respective $\mathcal{C}_{\mu_i}$ towards their destination with $i=1$ having arrived and transferred back to $\mathcal{C}_0$ before $t=2195$. Note that agent trajectories for $t \geq 2125$ are plotted.}
\label{fig:Avoidance Lapse}
\end{figure*}
A simulation was performed in MATLAB to verify the efficacy of the algorithm. Four agents are deployed to monitor the surface of an ellipsoid of revolution, $\mathcal{C}$, whose radius in the $xy$-plane is 80 and whose radius in the $z$-plane is 20. For each agent, $R=10$, $\sr_i=1$, $\alpha_i=30^{\circ}$, $k_u=1$, $k_v=5$, $k_w=1$, $k_r=0.1$, $k_s=0.1$, $\bar{r}_i=0.4$, $\bar{s}_i=0.4$, $U_{max}^{agt}=6$, and $T^{\star}=792$. Upon initialization of the simulation, $\mathcal{C}$ was set to a fully covered level of $C^{\star}=20$ which would begin decaying upon detection of the first intruder $k \in \{1,...,4\}$ at $t=600$ seconds. The four agents were deployed from $\mathcal{F}$ sequentially at times $t=0$, $t=\frac{T^{\star}}{4}$, $t=\frac{2T^{\star}}{4}$, and $t=\frac{T^{\star}}{4}$ seconds respectively. Upon deployment, each agent was initialized in local coverage mode with $\Phi_i=0$, $\Theta_i=\frac{\pi}{2}$, and $\Psi_i=0$. Intruders traveled in random directions with $U_{max}^{int}=0.7$, though were still constrained to always impact the surface, and were generated every 35 seconds beginning at $t=\frac{3T^{\star}}{4}$ seconds. The detection system had a lower bound on range $R_{det}=80$, decay rate parameter $\lambda_k=0.05$, and measurement variances $\sigma_\rho^2=0.0625$, $\sigma_\theta^2=0.25 \textnormal{ deg}^2$, and $\sigma_\psi^2=0.25 \textnormal{ deg}^2$ respectively
\begin{figure}[h]
\centering
\includegraphics[width=\columnwidth,clip]{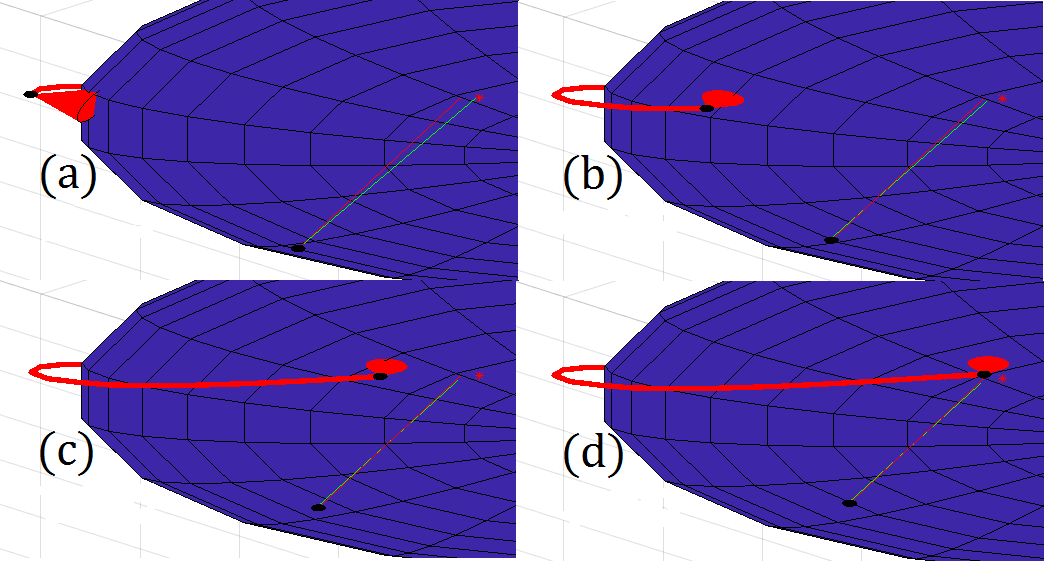}
\caption{Agent $i=2$ follows its geodesic trajectory to the predicted impact point of particle $k$ over time lapse (a)-(d). The true trajectory of the particle is indicated in red and the estimated trajectory in green.}
\label{fig:Particle_Intercept}
\end{figure}
Agents were able to successfully intercept all particles along their geodesic trajectories while actively avoiding collision over the entire duration of the attack (see Fig. \ref{fig:Avoidance Lapse} and Fig. \ref{fig:Particle_Intercept}); however, it should be noted that one avoidance anomaly occurred before the initial intruder was generated during the interval of $t=418-420$. As our sensing range for agents was $R=10$ and our simulation time step size was 1, it is clear that this anomaly occurred due to a selection of $U_{max}^{agt}$ that was too large relative to the time step. In a continuous time implementation, a transition to surface transfer mode would have occurred between $t=418$ and $t=419$ thus preventing collision. Aside from this anomaly, the simulation parameters adequately approximated the continuous time agent kinematics.\vspace{-2mm}

\begin{figure}[h]
\centering
\includegraphics[width=\columnwidth,clip]{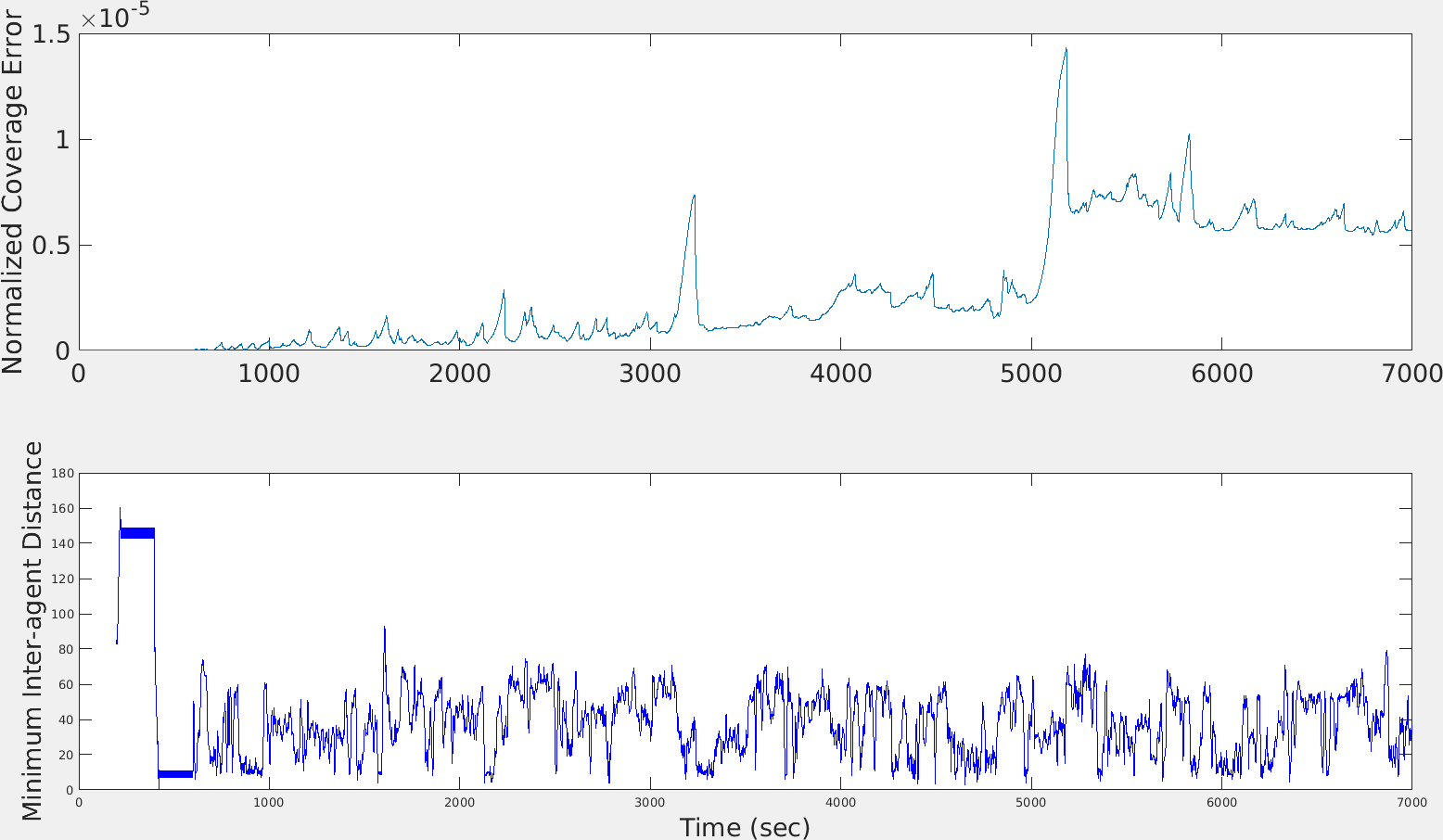}
\caption{The coverage error, normalized to the maximum possible value, is presented. Anomalies are observed at $t=3225$ and $t=5180$ respectively. The minimum distance between any two agents at any given time is also presented with an anomaly observed at $t=419$.}
\label{fig:Full_Trial}
\end{figure}

The coverage error on $\mathcal{C}$, normalized with respect to the maximum error in which all of $\mathcal{C}$ takes a value of zero for $Q\left(t,\tilde{p}\right)$, as well as the minimum inter-agent distance over time are presented in Fig. \ref{fig:Full_Trial}. The error tends to spike upon particle detections with agents effectively curtailing these spikes as they cover around the vicinity of predicted impact points in local coverage mode. Two particularly large spikes occur at $t=3225$ and $t=5180$ respectively. These anomalies are each associated with particle impacts occurring close to the equator of the ellipsoid where even small values of $\sigma_\theta^2$ and $\sigma_\psi^2$ result in an estimated particle trajectory that does not initially intersect $\mathcal{C}$ thus delaying an agent assignment. In both cases, the estimated trajectory did eventually intersect $\mathcal{C}$ with enough time to allow for agent interception. However, this delay in assignment significantly reduced the time the agent spent exploring in the vicinity of the predicted impact point thus contributing to a noticeable rise in the coverage error. One potential solution to this problem would be to prescribe some boundary tolerance to our surface $\mathcal{C}$ thus loosening our definition of an impacting particle for the sake of measurement uncertainty.

\begin{figure}[h]
\centering
\includegraphics[width=\columnwidth]{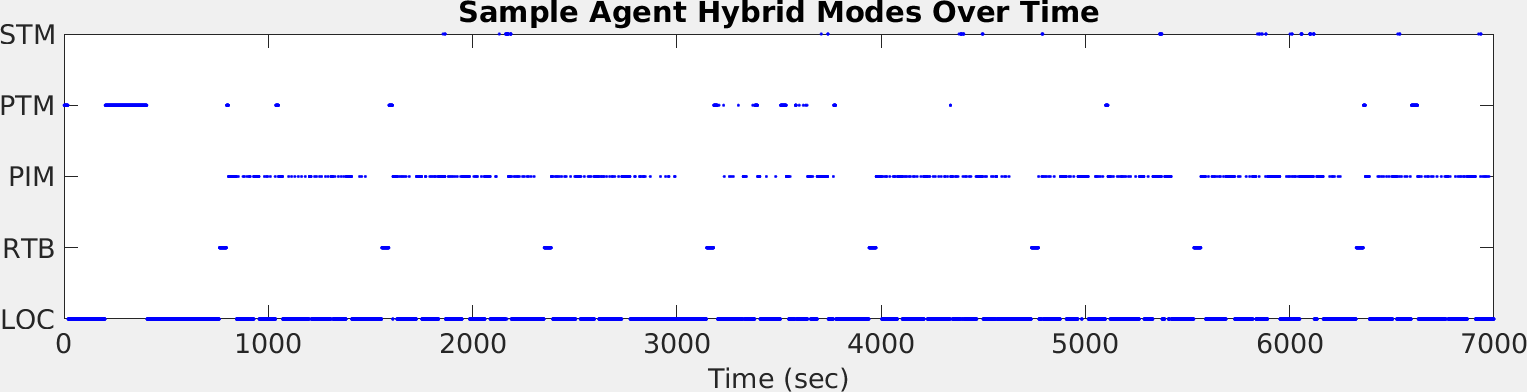}
\caption{A typical agent's hybrid modes are presented over time. Abbreviations from top to bottom refer to surface transfer mode, partition transfer mode, particle intercept mode, return to base, and local coverage mode respectively.}
\label{fig:i1i2hybrid}
\end{figure}
Agent $i=1$'s operating modes with respect to time are presented in Fig. \ref{fig:i1i2hybrid}. It should be noted that the most frequent transition out of particle intercept mode is to local coverage mode. This corresponds to an agent arriving at the estimated impact point of a particle and then surveying the local area up until the moment of impact. As the agent surveys it tends to hit the $\varepsilon_1$ proximity boundary to the impact point thus requiring a short operation in particle intercept mode to direct the agent back within the $\varepsilon_1$ boundary.

To demonstrate scalability, we have made an additional simulation with 100 agents available online at: \url{https://1drv.ms/f/s!AsiVOlIEkwNEgX2o1eV2hJ_bbaQU}.
\vspace{-3mm}\section{Conclusions}\vspace{-3mm} 
In this paper, we presented a hybrid formulation for the persistent coverage problem in an environment subject to stochastic intruders. This formulation was motivated in part by extravehicular applications of the NASA Mini AERCam. Agents operated with finite power resources and were required to periodically return to a refueling station while patrolling assigned latitude partitions along the surface of an ellipsoid. Formal guarantees were established on the ability of agents to intercept all intruders and the efficacy of the algorithm was demonstrated in simulation. This approach succeeds our previous work in \citet{Bentz_CDC_2017} and \citet{Bentz_ACC_2018} by extending the guarantees on intruder interception to an arbitrary number of collision avoidance maneuvers. It also removes singularities in the sensing function definition and redefines the guard conditions in a manner that supports a more effective use of local coverage around the vicinity of intruder impact points.
\section{Appendix}
\subsection{Additional Derivations for Local Coverage Strategy}
\begin{figure*}[!t]
\normalsize
\setcounter{mytempeqncnt}{\value{equation}}
\setcounter{equation}{26}\tiny
Expand $\frac{d}{dt}(S_{i}(\tilde{q}_{i}(t),\tilde{p}))$:
\vspace{-3mm}
\begin{align}
\begin{split}
\label{eq:10}
{}& \frac{d}{dt}(S_{i}(\tilde{q}_{i}(t),\tilde{p})) = \frac{\partial S_{i}}{\partial x_{i}}\dot{x}_{i}(t)+\frac{\partial S_{i}}{\partial y_{i}}\dot{y}_{i}(t)
+\frac{\partial S_{i}}{\partial z_{i}}\dot{z}_{i}(t)+\frac{\partial S_{i}}{\partial \Psi_{i}}\dot{\Psi}_{i}(t)+\frac{\partial S_{i}}{\partial \Theta_{i}}\dot{\Theta}_{i}(t)
=\left(\frac{\partial S_{i}}{\partial x_{i}}\cos\Theta\cos\Psi+\frac{\partial S_{i}}{\partial y_{i}}\cos\Theta\sin\Psi-\frac{\partial S_{i}}{\partial z_{i}}\sin\Theta\right)u_{i}(t) \\
&+\left(\frac{\partial S_{i}}{\partial x_{i}}(\sin\Phi\sin\Theta\cos\Psi-\cos\Phi\sin\Psi)\right.
\left.+\frac{\partial S_{i}}{\partial y_{i}}(\sin\Phi\sin\Theta\sin\Psi+\cos\Phi\cos\Psi)\right.
\left. +\frac{\partial S_{i}}{\partial z_{i}}\sin\Phi\cos\Theta\right)v_{i}(t) \\
&+\left(\frac{\partial S_{i}}{\partial x_{i}}(\cos\Phi\sin\Theta\cos\Psi+\sin\Phi\sin\Psi)\right.
\left.+\frac{\partial S_{i}}{\partial y_{i}}(\cos\Phi\sin\Theta\sin\Psi-\sin\Phi\cos\Psi)\right.
\left.+\frac{\partial S_{i}}{\partial z_{i}}\cos\Phi\cos\Theta\right)w_{i}(t) \\
&+\left(\frac{\partial S_{i}}{\partial \Psi_{i}}\sin\Phi\sec\Theta+\frac{\partial S_{i}}{\partial \Theta_{i}}\cos\Phi\right)r_{i}(t)
+\left(\frac{\partial S_{i}}{\partial \Psi_{i}}\cos\Phi\sec\Theta-\frac{\partial S_{i}}{\partial \Theta_{i}}\sin\Phi\right)s_{i}(t). 
\end{split} 
 \end{align}
Now introduce the following definitions:
\vspace{-3mm}
\begin{align} \label{eq:11}
a_{i0}(t,Q(t,\tilde{p}))&= \int\limits_{D_i} \-h''(C^{\star}C\left(\tilde{p}\right)-Q(t,\tilde{p}))S_{i}(\tilde{q}_{i}(t),\tilde{p})C\left(\tilde{p}\right)\frac{\partial Q(t,\tilde{p})}{\partial t}d\tilde{p}, \\
\label{eq:12}
a_{i1}(t,Q(t,\tilde{p}))&=\int\limits_{D_i} \-h'(C^{\star}C\left(\tilde{p}\right)-Q(t,\tilde{p}))C\left(\tilde{p}\right) \left(\frac{\partial S_{i}}{\partial x_{i}}\cos\Theta\cos\Psi\right.
\left.+\frac{\partial S_{i}}{\partial y_{i}}\cos\Theta\sin\Psi-\frac{\partial S_{i}}{\partial z_{i}}\sin\Theta\right)d\tilde{p}, \\
\label{eq:13}
a_{i2}(t,Q(t,\tilde{p}))&=\int\limits_{D_i} \-h'(C^{\star}C\left(\tilde{p}\right)-Q(t,\tilde{p}))C\left(\tilde{p}\right) \left(\frac{\partial S_{i}}{\partial x_{i}}(\sin\Phi\sin\Theta\cos\Psi\right.
\left.-\cos\Phi\sin\Psi)+\frac{\partial S_{i}}{\partial y_{i}}(\sin\Phi\sin\Theta\sin\Psi+\cos\Phi\cos\Psi)\right.
\left.+\frac{\partial S_{i}}{\partial z_{i}}\sin\Phi\cos\Theta\right)d\tilde{p}, \\
\label{eq:14}
a_{i3}(t,Q(t,\tilde{p}))&=\int\limits_{D_i} \-h'(C^{\star}C\left(\tilde{p}\right)-Q(t,\tilde{p}))C\left(\tilde{p}\right) \left(\frac{\partial S_{i}}{\partial x_{i}}(\cos\Phi\sin\Theta\cos\Psi+\sin\Phi\sin\Psi)\right.
\left.+\frac{\partial S_{i}}{\partial y_{i}}(\cos\Phi\sin\Theta\sin\Psi-\sin\Phi\cos\Psi)+\frac{\partial S_{i}}{\partial z_{i}}\cos\Phi\cos\Theta\right)d\tilde{p}, \\
\label{eq:15}
a_{i4}(t,Q(t,\tilde{p}))&=\int\limits_{D_i} \-h'(C^{\star}C\left(\tilde{p}\right)-Q(t,\tilde{p}))C\left(\tilde{p}\right) \left(\frac{\partial S_{i}}{\partial \Psi_{i}}\sin\Phi\sec\Theta+\frac{\partial S_{i}}{\partial \Theta_{i}}\cos\Phi\right)d\tilde{p}, \\
\label{eq:16}
a_{i5}(t,Q(t,\tilde{p}))&=\int\limits_{D_i} \-h'(C^{\star}C\left(\tilde{p}\right)-Q(t,\tilde{p}))C\left(\tilde{p}\right) \left(\frac{\partial S_{i}}{\partial \Psi_{i}}\cos\Phi\sec\Theta-\frac{\partial S_{i}}{\partial \Theta_{i}}\sin\Phi\right)d\tilde{p}.
\end{align}
One can then rewrite \eqref{eq:9} as:
\vspace{-3mm}
\begin{align} \label{eq:17}
\dot{\hat{e}}_{i}(t)=a_{i0}(t,Q(t,\tilde{p}))-u_{i}(t)a_{i1}(t,Q(t,\tilde{p}))-v_{i}(t)a_{i2}(t,Q(t,\tilde{p}))-w_{i}(t)a_{i3}(t,Q(t,\tilde{p}))
-r_{i}(t)a_{i4}(t,Q(t,\tilde{p}))-s_{i}(t)a_{i5}(t,Q(t,\tilde{p})).
\end{align}
\normalsize
\setcounter{equation}{\value{mytempeqncnt}}
\hrulefill
\vspace*{4pt}
\end{figure*}
Our local coverage control laws are derived via differentiation of \eqref{Global Coverage Error} of which we seek to reduce the rate of growth. It is a volume integral, so a few mathematical preliminaries are required. Recall the generalized transport theorem (GTT) \citep{Slattery}: $\frac{d}{dt}\int\limits_{R_{(s)}}f dV =\int\limits_{R_{(s)}} \frac{\partial f}{\partial t}\,dV + 
\int\limits_{S_{(s)}} f \mathbf{v_{(s)}} \cdot \mathbf{n}\,dA$, where $f$ is any scalar-, vector-, or tensor-valued function of position and time, $S_{(s)}$ is the boundary of the volume $R_{(s)}$ over which $f$ is integrated, $\mathbf{n}$ is the unit vector normal to the boundary, and $\mathbf{v_{(s)}}$ is the velocity of the boundary. $V$ and $A$ refer to volume and area respectively. Invoking GTT allows for differentiation of \eqref{Global Coverage Error} with respect to time:\vspace{-3mm}
\begin{align} \label{E_dot}
\begin{split}
\dot{E}(t) = &\int\limits_{\mathcal{D}} \ h'(C^{\star}C\left(\tilde{p}\right)-Q(t,\tilde{p})) \left( \frac{- \partial Q(t,\tilde{p})}{\partial t}\right) \,d\tilde{p}\\
&+ \int\limits_{\partial \mathcal{D}}\big(h(C^{\star}C\left(\tilde{p}\right)-Q(t,\tilde{p})) \big) \mathbf{v_{(s)}} \cdot \mathbf{n}\,dA,
\end{split}
\end{align}
where $\partial \mathcal{D}$ is the boundary of $\mathcal{D}$. $\mathcal{D}$ is time invariant and thus $\mathbf{v_{(s)}}=0$. \eqref{E_dot} reduces to $\dot{E}(t)= \int\limits_{D} \ h'(C^{\star}C\left(\tilde{p}\right)-Q(t,\tilde{p})) \left( \frac{- \partial Q(t,\tilde{p})}{\partial t}\right) \,d\tilde{p}$, which expands to:\vspace{-3mm}
\scriptsize
\begin{align}
\begin{split}
\label{E_dot expanded}
\dot{E}(t) &=- \int\limits_{D} \ h'(C^{\star}C\left(\tilde{p}\right)-Q(t,\tilde{p})) \biggl(\sum_{i=1}^{N}S_{i}(\tilde{q}_{i}(t),\tilde{p})C\left(\tilde{p}\right) \\
&-\sum_{k=1}^{N_p}\Lambda_k\left(t,\tilde{p}\right)C\left(\tilde{p}\right)\biggr) d\tilde{p} \\
          &=\sum_{i=1}^{N}\underbrace{\int\limits_{D} \ -h'(C^{\star}C\left(\tilde{p}\right)-Q(t,\tilde{p})) S_{i}(\tilde{q}_{i}(t),\tilde{p})C\left(\tilde{p}\right) d\tilde{p}}_{=\hat{e}_i(t)} \\
&-\sum_{k=1}^{N_p}\underbrace{\int\limits_{D} \ -h'(C^{\star}C\left(\tilde{p}\right)-Q(t,\tilde{p}))\Lambda_k\left(t,\tilde{p}\right)C\left(\tilde{p}\right) d\tilde{p}}_{=\tilde{e}_k (t)}    \\    
&=\sum_{i=1}^{N}\hat{e}_i (t)-\sum_{k=1}^{N_p}\tilde{e}_k (t).
\end{split}
\end{align}
\normalsize
$\hat{e}_i (t)$ is the rate of change of the coverage error due to the motion of the agents while $\tilde{e}_k (t)$ is the rate of change of the coverage error due to a contrived information decay surrounding the predicted impact point of particle $k$ on $\mathcal{C}$. Our strategy is to control the agents' kinematics, recovered in the derivative of $\hat{e}_i (t)$, to decrease \eqref{E_dot expanded}. Note that we do not presume that our local coverage strategy provides any additional bounds on \eqref{Global Coverage Error}. Nor do we provide guarantees on the rate of growth of this contrived quantity. Curtailing the growth of the coverage error simply imparts the desired effect of active exploration in the vicinity of impact points into our system. Using this strategy, the agents actively seek to increase their rate of coverage by rotating and/or translating $\mathcal{S}_i$ to be encompass the most uncovered space in the local vicinity.

Taking the derivative of $\hat{e}_i (t)$ with respect to time yields: \vspace{-3mm}
\small
\begin{multline}\label{eq:9} 
\dot{\hat{e}}_{i}(t)=\int\limits_{D_i} \ \biggl(h''(C^{\star}C\left(\tilde{p}\right)-Q(t,\tilde{p}))S_{i}(\tilde{q}_{i}(t),\tilde{p})C\left(\tilde{p}\right)\frac{\partial Q(t,\tilde{p})}{\partial t} \\
-h'(C^{\star}C\left(\tilde{p}\right)-Q(t,\tilde{p})) \frac{d}{dt}(S_{i}(\tilde{q}_{i}(t),\tilde{p}))C\left(\tilde{p}\right)\biggr)d\tilde{p}.
\end{multline}
\normalsize
The sensing footprint is independent of $\Phi_{i}$ assuming that the centerline of the spherical sector is aligned with the $\hat{x}_{\mathcal{B}_i}$ axis. $\frac{d}{dt}(S_{i}(\tilde{q}_{i}(t),\tilde{p}))$ is expanded in \eqref{eq:10} and through the definitions in (28-33) one may restate \eqref{eq:9} as \eqref{eq:17}.
\addtocounter{equation}{8}
If one were to command zero inputs to this system, it becomes clear that $a_{i0}(t,Q(t,\tilde{p}))$ may be physically interpreted as the rate at which the coverage rate is reducing due to information saturation at any particular position and orientation of the sensing footprint, $\mathcal{S}_i$. As the footprint remains stationary, there are diminishing returns on the value of newly acquired information. Thus, the additional terms in \eqref{eq:17} allow for the coverage rate to be increased by mobilizing the sensor. One strategy is that of \eqref{coverage control law}.

\subsection{Formal Hybrid Formulation}
To provide a compact notation in this section, define $\acute{f}_i=\frac{x_i^2}{\left(x_{\mathcal{C},r}+\sr_i\right)^2}+\frac{y_i^2}{\left(y_{\mathcal{C},r}+\sr_i\right)^2}+\frac{z_i^2}{\left(z_{\mathcal{C},r}+\sr_i\right)^2}$. The coverage strategy for agent $i$ is represented by the hybrid automaton in Fig. \ref{fig:Automaton}, described by the following entities \citep{Lygeros}:
\begin{itemize}
\item A set of discrete states: $Z_{i}=\{\zeta_{i0},\zeta_{i1},\zeta_{i2},\zeta_{i3},\zeta_{i4}\},$
\item A set of continuous states:
$\tilde{q}_i=\{x_i,y_i,z_i,\Phi_i,\Theta_i,\Psi_i\},$
\item A vector field:
$\newline f(\zeta_{i0},\tilde{q}_i)  =\mathcal{R}
\left[
u_i^{loc} \;
v_i^{loc} \;
w_i^{loc} \;
0 \;
r_i^{loc} \;
s_i^{loc} \;
\right]^{T},
\newline f(\zeta_{i1},\tilde{q}_i)  = \mathcal{R}
\left[
u_{i}^{rtb} \;
v_{i}^{rtb} \;
w_{i}^{rtb} \;
q_{i}^{rtb} \;
r_{i}^{rtb} \;
s_{i}^{rtb} \;
\right]^{T},
\newline f(\zeta_{i2},\tilde{q}_i)  =\mathcal{R}
\left[
u_i^{pim} \;
v_i^{pim} \;
w_i^{pim} \;
q_{i}^{pim} \;
r_{i}^{pim} \;
s_{i}^{pim} \;
\right]^{T},
\newline f(\zeta_{i3},\tilde{q}_i)  =\mathcal{R}
\left[
u_i^{ptm} \;
v_i^{ptm} \;
w_i^{ptm} \;
q_{i}^{ptm} \;
r_{i}^{ptm} \;
s_{i}^{ptm} \;
\right]^{T},
\newline f(\zeta_{i4},\tilde{q}_i)  =\mathcal{R}
\left[
u_i^{stm} \;
v_i^{stm} \;
w_i^{stm} \;
0 \;
0 \;
0 \;
\right]^{T}
\textnormal{where} \; \mathcal{R}= \begin{bmatrix} 
\mathcal{R}_1 & 0 \\
0 & \mathcal{R}_2
\end{bmatrix},$
\item A set of initial states:
$\{\zeta_{i3}\} \times \{\tilde{q}_{i} \in \mathbb{R}^6 \mid p_i=\mathcal{F} \\
\land \, \Phi_{i} \in \left[-\pi,+\pi\right] \, \land \, \Theta_{i} \in \left[\frac{-\pi}{2},\frac{+\pi}{2}\right] \, \land \, \Psi_{i} \in \left[-\pi,+\pi\right]\},$
\item A domain: $
Dom \left(\zeta_{i0}\right)=\{\tilde{q}_i \in \mathbb{R}^6 \mid \acute{f}_i \geq 1 \land \\ \left(i_p \in \{2,...,N\} \implies \bar{z}_{{i_p}-1} \leq z_i \leq \bar{z}_{{i_p}-2}\right)\},
\\
Dom \left(\zeta_{i1}\right)=\{\tilde{q}_i \in \mathbb{R}^6 \mid \acute{f}_i \geq 1\},
\\
Dom \left(\zeta_{i2}\right)=\{\tilde{q}_i \in \mathbb{R}^6 \mid \acute{f}_i \geq 1\},
\\
Dom \left(\zeta_{i3}\right)=\{\tilde{q}_i \in \mathbb{R}^6 \mid \acute{f}_i \geq 1 \land \\ \left(i_p \in \{2,...,N\} \implies z_i<\bar{z}_{{i_p}-1} \lor z_i > \bar{z}_{{i_p}-2}\right)\},
\\
Dom \left(\zeta_{i4}\right)=\{\tilde{q}_i \in \mathbb{R}^6 \mid \acute{f}_i \geq 1\},
$
\item A set of edges: $E=\{\left(\zeta_{i0},\zeta_{i1}\right),\left(\zeta_{i0},\zeta_{i2}\right), \left(\zeta_{i0},\zeta_{i3}\right), \\ \left(\zeta_{i0},\zeta_{i4}\right), \left(\zeta_{i1},\zeta_{i3}\right),\left(\zeta_{i2},\zeta_{i0}\right),\left(\zeta_{i2},\zeta_{i1}\right),\left(\zeta_{i2},\zeta_{i3}\right),\\
\left(\zeta_{i2},\zeta_{i4}\right),
\left(\zeta_{i3},\zeta_{i0}\right),\left(\zeta_{i3},\zeta_{i2}\right),
\left(\zeta_{i3},\zeta_{i4}\right), \left(\zeta_{i4},\zeta_{i0}\right), \\ \left(\zeta_{i4},\zeta_{i2}\right),\left(\zeta_{i4},\zeta_{i3}\right),
\},$
\item A set of guard conditions:\scriptsize $\\ G\left(\zeta_{i0},\zeta_{i1}\right) = \{i_p=1 \land t_{i\mathcal{F}} \geq T^{\star}-\frac{\pi g_{\mathcal{C}_0}}{2U_{max}^{agt}}\left(x_{\mathcal{C}_0,r}+z_{\mathcal{C}_0,r}\right)\}, \\ G\left(\zeta_{i0},\zeta_{i2}\right) = \{\left(\exists k \mid i=i_k\right) \land \left(\lVert p_i-\vectorproj[\mathcal{C}_{\mu_i}]{p_{id}}  \rVert > \varepsilon_1\right)\}, \\
G\left(\zeta_{i0},\zeta_{i3}\right) =\{i_p \neq 1 \land \left(z_i<\bar{z}_{{i_p}-1} \lor z_i > \bar{z}_{{i_p}-2}\right)\},\\
G\left(\zeta_{i0},\zeta_{i4}\right)=\{\lVert p_i-p_j \rVert \leq R \land i_{pr} \neq \argmax_{\tilde{j}}\left(t_{\tilde{j}\mathcal{F}}\right)\},\\
G\left(\zeta_{i1},\zeta_{i3}\right) =\{\lVert p_i-\mathcal{F} \rVert \leq \varepsilon_1 \land t_{i\mathcal{F}}=T^{\star}\}, \\
G\left(\zeta_{i2},\zeta_{i0}\right)=\{\bigl(t \geq t_{ck} \land \\ \left. \bigl(\left(i_p \in \{2,...,N\} \land \bar{z}_{{i_p}-1} \leq z_i \leq \bar{z}_{{i_p}-2}\right)
\lor \right. \\ \left. \left. \left(i_p=1 \land t_{i\mathcal{F}} < T^{\star}-\frac{\pi g_{\mathcal{C}_0}}{2U_{max}^{agt}}\left(x_{\mathcal{C}_0,r}+z_{\mathcal{C}_0,r}\right)                  \right)\right)\right) \lor \\ \left(t<t_{ck} \land \lVert p_i-\vectorproj[\mathcal{C}_{\mu_i}]{p_{id}}  \rVert \leq \varepsilon_1 \land \mu_i=0\right)\},\\
G\left(\zeta_{i2},\zeta_{i1}\right) = \{\lVert p_i- \vectorproj[\mathcal{C}_{\mu_i}]{p_{id}} \rVert \leq \varepsilon_1 \land \mu_i=0 \land t \geq t_{ck} \land i_p=1 \land t_{i\mathcal{F}} \geq T^{\star}-\frac{\pi g_{\mathcal{C}_0}}{2U_{max}^{agt}}\left(x_{\mathcal{C}_0,r}+z_{\mathcal{C}_0,r}\right)\} \\
G\left(\zeta_{i2},\zeta_{i3}\right) = \{\lVert p_i-\vectorproj[\mathcal{C}_{\mu_i}]{p_{id}} \rVert \leq \varepsilon_1 \land \mu_i=0 \land t \geq t_{ck} \land i_p \in \{2,...,N\} \land \left(z_i<\bar{z}_{{i_p}-1} \lor z_i> \bar{z}_{{i_p}-2}\right)\}, \\
G\left(\zeta_{i2},\zeta_{i4}\right) =G\left(\zeta_{i0},\zeta_{i4}\right) \lor \{\lVert p_i-\vectorproj[\mathcal{C}_{\mu_i}]{p_{id}} \rVert \leq \varepsilon_1 \land \left(\lVert p_i-p_j \rVert > R, \forall j \lor i_{pr} = \argmax_{\tilde{j}}\left(t_{\tilde{j}\mathcal{F}}\right)\right) \land \mu_i>0\}, \\
G\left(\zeta_{i3},\zeta_{i0}\right) = \{i_p=1 \lor \left(i_p \neq 1 \land \bar{z}_{{i_p}-1} \leq z_i \leq \bar{z}_{{i_p}-2}\right)\},\\
G\left(\zeta_{i3},\zeta_{i2}\right) = G\left(\zeta_{i0},\zeta_{i2}\right), \\
G\left(\zeta_{i3},\zeta_{i4}\right) = G\left(\zeta_{i0},\zeta_{i4}\right)\lor \{\lVert p_i-\vectorproj[\mathcal{C}_{\mu_i}]{p_{id}} \rVert \leq \varepsilon_1 \land \left(\lVert p_i-p_j \rVert > R, \forall j \lor i_{pr} = \argmax_{\tilde{j}}\left(t_{\tilde{j}\mathcal{F}}\right)\right) \land \mu_i>0\}, \\
G\left(\zeta_{i4},\zeta_{i0}\right)=\{f_i=0 \land \lvert\ln{\left(\frac{\lVert n_i \rVert - \sr_i}{\left(\gamma+\mu_i\right)R-\sr_i}\right)}\rvert < \varepsilon_2 \land \left(i_p=1 \lor \\ \left(i_p \in \{2,...,N\} \land \bar{z}_{{i_p}-1} \leq z_i \leq \bar{z}_{{i_p}-2}\right)\right)\}, \\
G\left(\zeta_{i4},\zeta_{i2}\right) = \{f_i=1 \land \lvert\ln{\left(\frac{\lVert n_i \rVert - \sr_i}{\left(\gamma+\mu_i\right)R-\sr_i}\right)}\rvert < \varepsilon_2\}, \\
G\left(\zeta_{i4},\zeta_{i3}\right) = \{f_i=0 \land \lvert\ln{\left(\frac{\lVert n_i \rVert - \sr_i}{\left(\gamma+\mu_i\right)R-\sr_i}\right)}\rvert < \varepsilon_2 \land \left(i_p \neq 1 \land \left(z_i<\bar{z}_{{i_p}-1} \lor z_i > \bar{z}_{{i_p}-2}\right)\right)\}.$
\item Additional parameters include a clock set: $C=\{t_{i \mathcal{F}}\},$ a flag: $f_i \in \{0,1\}$, an assignment index $\mu_i=\{0,...,N-1\}$ and,
\item A reset map: $
R\left(\zeta_{i0},\zeta_{i2},f_i\right) =\{1\},
R\left(\zeta_{i0},\zeta_{i4},\mu_i\right)= \\ \{\mu_i+1\},
R\left(\zeta_{i1},\zeta_{i3},t_{i \mathcal{F}}\right) = \{0\},
R\left(\zeta_{i2},\zeta_{i0},f_i\right) = \\ \{0 \textnormal{ if } t \geq t_{ck}; 1 \textnormal{ otherwise}\},
R\left(\zeta_{i2},\zeta_{i1},f_i\right) =\{0\}, \\
R\left(\zeta_{i2},\zeta_{i3},f_i\right) =\{0\},
R\left(\zeta_{i2},\zeta_{i4},\mu_i\right)=\{0 \textnormal{ if } \lVert p_i-\vectorproj[\mathcal{C}_{\mu_i}]{p_{id}} \rVert \\ \leq \varepsilon_1 \land \left(\lVert p_i-p_j \rVert > R, \forall j \lor i_{pr} = \argmax_{\tilde{j}}\left(t_{\tilde{j}\mathcal{F}}\right)\right) \land \mu_i>0; \mu_i+1 \textnormal{ otherwise}\},
R\left(\zeta_{i3},\zeta_{i2},f_i\right) =\{1\}, \\
R\left(\zeta_{i3},\zeta_{i4},\mu_i\right)=\{0 \textnormal{ if } \lVert p_i-\vectorproj[\mathcal{C}_{\mu_i}]{p_{id}} \rVert \leq \varepsilon_1 \land \\ \left(\lVert p_i-p_j \rVert > R, \forall j \lor i_{pr} = \argmax_{\tilde{j}}\left(t_{\tilde{j}\mathcal{F}}\right)\right) \land \mu_i>0; \\ \mu_i+1 \textnormal{ otherwise}\},
R\left(\zeta_{i4},\zeta_{i0},\mu_i\right)=\{0\},$
and continuous states do not reset between transitions.
\end{itemize}
\vspace{-2mm}
\bibliography{Bentz_Cov_References}

\end{document}